\documentclass{article}
\usepackage{graphicx}  
\usepackage{amsmath}   
\usepackage[compress]{cite}
\usepackage{amssymb}   
\usepackage{bm} 
\usepackage{dcolumn}
\usepackage{color}
\usepackage{mathrsfs}
\usepackage{amsfonts}
\usepackage{varioref}
\usepackage{float}
\restylefloat{table}
\usepackage[caption=false]{subfig}
\RequirePackage[colorlinks,citecolor=blue,urlcolor=magenta,linkcolor=blue]{hyperref}
\input epsf

\allowdisplaybreaks
\def\gr{general relativity}

\labelformat{section}{Section #1} 
\labelformat{subsection}{Section #1} 
\labelformat{subsubsection}{Section #1}
\labelformat{subsubsubsection}{Section #1}
\labelformat{equation}{Eq.~(#1)} 
\labelformat{figure}{Fig.~#1} 
\labelformat{subfigure}{Fig.~\thefigure#1} 
\labelformat{table}{Table~#1} 
\labelformat{appendix}{Appendix #1}
\definecolor{ao(english)}{rgb}{0.0, 0.5, 0.0}
\definecolor{brightgreen}{rgb}{0.4, 1.0, 0.0}
\definecolor{darkpastelgreen}{rgb}{0.01, 0.75, 0.24}
\definecolor{darkpastelblue}{rgb}{0.47, 0.62, 0.8}
\definecolor{darkpastelpurple}{rgb}{0.59, 0.44, 0.84}
\definecolor{flame}{rgb}{0.89, 0.35, 0.13}
\title{Signatures of Einstein-Maxwell dilaton-axion gravity from the observed jet power and the radiative efficiency}
\author{Indrani Banerjee\footnote{banerjeein@nitrkl.ac.in}~,
Bhaswati Mandal\footnote{tpbm3@iacs.res.in}~
and
and Soumitra SenGupta\footnote{tpssg@iacs.res.in}\\
{\small{\hspace{-1.5cm}School of Physical Sciences, Indian Association for the Cultivation of Science,
2A \& 2B Raja S. C. Mullick Road, Kolkata-700032, India}}\\
{\small{\hspace{-1.5cm} Department of Physics and Astronomy, National Institute of Technology, Rourkela, Odisha-769008, India }}}

\date{ } 
\begin{document}
\maketitle 
\begin{abstract}
The Einstein-Maxwell dilaton-axion (EMDA) gravity arises in the low energy effective action of
the heterotic string theory and provides a simple framework to explore the signatures of the same.
The dilaton and the axion fields inherited in the action from string compactifications have interesting
consequences in inflationary cosmology and in explaining the present accelerated expansion of the uni-
verse. It is therefore worthwhile to search for the footprints of these fields in the available astrophysical
observations. Since Einstein gravity is expected to receive quantum corrections in the high curvature
domain, the near horizon regime of black holes seems to be the ideal astrophysical laboratory to test
these deviations from general relativity. Exact, stationary and axisymmetric black hole solution in
EMDA gravity corresponds to the Kerr-Sen spacetime which carries dilaton charge, while the angular
momentum is sourced by the axion field. The ballistic jets and the peak emission of the continuum
spectrum from the accretion disk are believed to be launched very close to the event horizon and hence
should bear the imprints of the background spacetime. We compute the jet power and the radiative
efficiency derived from the continuum spectrum in the Kerr-Sen background and compare them with
the corresponding observations of microquasars. Our analysis reveals that Kerr black holes are more
favored compared to Kerr-Sen black holes with dilaton charges.

\end{abstract}

\section{Introduction}
\label{Intro}
General relativity is the most competent theory of gravity, till date, due to its unprecedented success in explaining a plethora of observations \cite{Will:2005yc,Will:1993ns,Will:2005va,Berti:2015itd} namely, the perihelion precession of mercury, the bending of light, the gravitational redshift of radiation from distant stars, to name a few. An accelerating universe, the existence of enigmatic objects like black holes and the detection of gravitational waves due to colliding black holes and neutron stars are some of the remarkable predictions of \gr\ which increasingly received observational confirmations with the advent of advanced ground-based and space-based missions \cite{Abbott:2017vtc,TheLIGOScientific:2016pea,Abbott:2016nmj,TheLIGOScientific:2016src,Abbott:2016blz}. The recent observation of the image of the black hole M87* by the Event Horizon Telescope Collaboration has further added to its phenomenal success \cite{Fish:2016jil,Akiyama:2019cqa,Akiyama:2019brx,Akiyama:2019sww,Akiyama:2019bqs,Akiyama:2019fyp,Akiyama:2019eap}. Yet, it is instructive to explore other alternate gravity models since \gr\ loses its predictive power at the black hole and the big-bang singularities \cite{Penrose:1964wq,Hawking:1976ra,Christodoulou:1991yfa} and the ultraviolet character of gravity continues to be ill-understood. On the observational front, one needs to invoke the exotic dark matter and the dark energy \cite{Bekenstein:1984tv,Perlmutter:1998np,Riess:1998cb} to explain the galactic rotation curves and the accelerated expansion of the universe respectively, if \gr\ is considered to be the correct theory of gravity. 

It is therefore believed that close to the Planck scale, \gr\ must receive corrections from a more complete theory of gravity which also incorporates its quantum character \cite{Rovelli:1996dv,Dowker:2005tz,Ashtekar:2006rx,Kothawala:2013maa}. A variety of alternate gravity models have therefore been put forward which can potentially fulfil the deficiencies in \gr. This includes, higher dimensional models \cite{Shiromizu:1999wj,Dadhich:2000am,Harko:2004ui,
Carames:2012gr,Kobayashi:2006jw,Chakraborty:2014xla,Chakraborty:2015bja}, higher curvature gravity, e.g., $f(R)$ \cite{Nojiri:2003ft,Nojiri:2006gh,Capozziello:2006dj}and Lanczos Lovelock models \cite{Lanczos:1932zz,Lanczos:1938sf,Lovelock:1971yv,Padmanabhan:2013xyr}, and the scalar-tensor theories of gravity \cite{Horndeski:1974wa,Sotiriou:2013qea,Babichev:2016rlq,Charmousis:2015txa,Bhattacharya:2016naa}. Many of these models are inspired from string theory \cite{Horava:1995qa,Horava:1996ma,Polchinski:1998rq,Polchinski:1998rr}
which provides a framework to unify all the known forces of nature under a single umbrella. The Einstein-Maxwell dilaton-axion (EMDA) gravity, which is central to this work is one such string inspired scalar-vector-tensor theory of gravity. Such a theory 
arises in the low energy effective action of superstring theories \cite{Sen:1992ua} when the ten dimensional heterotic string theory is compactified on a six dimensional torus $T^6$. The resultant four dimensional action comprises of $N=4$, $d=4$ supergravity coupled to $N=4$ super Yang-Mills theory, in the low-energy limit. By introducing equal numbers of Kaluza-Klein and winding number modes for each cycle, this effective action can be further truncated to a pure supergravity theory exhibiting $S$ and $T$ dualities. The bosonic sector of this $N=4$, $d=4$ supergravity with a vector field is known as the Einstein-Maxwell dilaton-axion (EMDA) gravity \cite{Rogatko:2002qe} which provides a simple framework to study classical solutions. 

The EMDA theory of gravity comprises of the scalar field dilaton and the pseudo scalar axion coupled to the metric and the Maxwell field. The dilaton and the axion fields owe their origin from string compactifications and have interesting consequences in inflationary cosmology and late time acceleration of the universe \cite{Sonner:2006yn,Catena:2007jf}.
Various classes of black hole solutions of string inspired low-energy effective theories have been constructed \cite{Gibbons:1987ps,Garfinkle:1990qj,Horowitz:1991cd,Kallosh:1993yg} which differ significantly from the general relativistic scenario. While the static and spherically symmetric black hole solutions bear non-trivial charges associated with the dilaton and the anti-symmetric tensor gauge fields, the charge neutral rotating solution  in string theory is identical with the Kerr metric in \gr\  \cite{Thorne:1986iy}. 
However, the stationary and axi-symmetric black hole solution in EMDA gravity represents the charged, rotating Kerr-Sen metric which is similar but not identical to the Kerr-Newman solution in \gr. Although the Kerr-Sen spacetime bears strong resemblance with the Kerr-Newman background, the inherent geometry of the two black holes vary significantly. The distinguishing properties of the two spacetimes have been extensively studied \cite{Hioki:2008zw,Pradhan:2015yea,Guo:2019lur,Uniyal:2017yll} in the past.
Since string theory incorporates the quantum nature of gravity and provides a promising framework for force unification, it is instructive to study the observational features of the Kerr-Sen metric, which in turn can provide an indirect testbed for string theory. 
Astrophysical implications of the Kerr-Sen black hole have been explored extensively \cite{Gyulchev:2006zg,An:2017hby,Younsi:2016azx,Hioki:2008zw,Mizuno:2018lxz} which includes study of null geodesics, photon motion, strong gravitational lensing and black hole shadow. 
Rotation of the polarization of circularly
polarized light in the vicinity of the Kerr-Sen black hole and an observation of a shadow of the same has been studied in \cite{Narang:2020bgo}. 

Astrophysical systems containing black holes are known to exhibit energetic transient jets \cite{Mirabel:1994rb,Fender:2004aw} which are often believed to be powered by the rotational energy of the black hole through the magnetic fields generated in the surrounding accreting plasma \cite{Blandford:1977ds}. 
Jets are believed to be launched very close to the event horizon and consequently the related jet power encodes the information of the underlying spacetime. Similarly, the continuum spectrum emitted from the accretion disk surrounding the black holes bears the imprints of the background metric and hence can be used to extract information about the same \cite{Novikov:1973kta}.
In this work we explore the role of the Kerr-Sen background in launching jets and in affecting the continuum emission from the accretion disk. We use the radiative efficiency derived from the peak emission of the continuum spectrum and the observed jet power of microquasars as the observables. This in turn enables us to constrain the metric parameters and hence provide an indirect observational evidence of string theory.

The paper is organized as follows: In \ref{S2} we provide a brief overview of the Einstein-Maxwell dilaton-axion (EMDA) gravity and the associated black hole solution. We explain the role of the background metric in affecting the radiative efficiency and the jet power in \ref{S3}. The theoretically computed jet power and radiative efficiency are compared with the corresponding observations of several black hole sources in \ref{S4}. Finally, we conclude with a summary of our findings and discussion of our results in \ref{Sec5}. 

We use (-,+,+,+) as the metric convention and will work with geometrized units taking $G=c=1$.


\section{Einstein-Maxwell-dilaton-axion gravity: An overview }
\label{S2}

The Einstein-Maxwell dilaton-axion (EMDA) gravity provides a generalization of the Einstein-Maxwell action comprising of the couplings between the metric $g_{\mu\nu}$, the $U(1)$ gauge field $A_\mu$, the dilaton field $\chi$ and the third rank anti-symmetric tensor field $\mathcal{H}_{\mu\nu\alpha}$. The resultant action upto $\mathcal{O}(\alpha^\prime)$ (where $\alpha^\prime$ is the inverse string tension) in the expansion of the effective action for the heterotic string theory is given by \cite{Campbell:1992hc}

\begin{align}
\label{S2-1}
\mathcal{S} = \int\sqrt{-g}d^{4}x\bigg{[}~ \frac{\mathcal{R}}{2\kappa^2} -\frac{1}{2} \partial_{\nu}\chi\partial^{\nu}\chi -6e^{-2\sqrt{2}\kappa \chi}\mathcal{H}_{\rho\sigma\delta}\mathcal{H}^{\rho\sigma\delta} -\frac{\alpha^\prime}{16\kappa^2} e^{-\sqrt{2}\kappa\chi}\mathcal{F}_{\alpha\beta}\mathcal{F}^{\alpha\beta}\bigg{]} 
\end{align}
where, $g$ is the determinant and $\mathcal{R}$ the Ricci scalar with respect to the 4-dimensional metric $g_{\mu\nu}$. The dilaton field is denoted by $\chi$ while $\mathcal{F}_{\mu\nu}$ represents the second rank antisymmetric Maxwell field strength tensor such that $\mathcal{F}_{\mu\nu}=\nabla_\mu A_\nu-\nabla_\nu A_\mu$, where $A_\mu$ is the $U(1)$ gauge field. In \ref{S2-1}, the three-rank antisymmetric tensor field $\mathcal{H}_{\rho\sigma\delta}$ is defined by the relation,
\begin{align}
\label{S2-2}
\mathcal{H}_{\rho\sigma\delta}=\nabla_{[\delta} B_{\rho\sigma]}-\frac{\alpha^\prime}{32\kappa\sqrt{-g}}A_{[\delta} F_{\rho\sigma]} 
\end{align}
where $B_{\mu\nu}$ is the second rank anti-symmetric tensor gauge field also known as the Kalb-Ramond field and the square brackets associated with the indices represent the cyclic sum over the indices. The cyclic permutation of $A_\mu$ and $F_{\mu\nu}$ in \ref{S2-2} represents the gauge Chern-Simons term. It is important to note that the action in \ref{S2-1} is the low energy effective action arising in heterotic string theory taken upto $\mathcal{O}(\alpha^\prime)$ and truncated to contain only those terms which involves upto a maximum of two derivatives. Therefore, higher curvature terms in \ref{S2-1} and Lorentz Chern Simons terms in the definition of \ref{S2-2} are not taken into account in this theory \cite{Campbell:1992hc,Sen:1992ua}. Such a truncation holds good when we are interested in regions far away from the spacetime singularity \cite{Campbell:1992hc,Sen:1994eb,Narang:2020bgo}. Moreover, it leads to exact black hole solutions in string theory which we will discuss towards the end of this section.
We further mention that the effective string action also contains several gauge fields, of which only the $U(1)$ component is kept in the action, since we will be eventually interested in black hole solutions carrying $U(1)$ charge \cite{Sen:1992ua,Campbell:1992hc}. 

In four dimensions $\mathcal{H}_{\mu\nu\alpha}$ can be written in terms of the pseudo-scalar field $\xi$, also known as the axion field \cite{Campbell:1992hc,Sen:1992ua,Narang:2020bgo}, such that,
\begin{align}
\label{S2-3}
\mathcal{H}_{\rho\sigma\delta} = \frac{e^{\sqrt{2}\kappa\chi}}{6\sqrt{2}}\epsilon_{\rho\sigma\delta\gamma}\partial^{\gamma}\xi-\frac{\alpha^\prime}{32\kappa\sqrt{-g}}A_{[\rho}F_{\sigma\delta]} 
\end{align}
In terms of the axion field the action in \ref{S2-1} can be written as,
\begin{align}
\label{S2-4}
\mathcal{S} = \int\bigg{[}\frac{\mathcal{R}}{2\kappa^2} - \frac{1}{2}\partial_{\nu}\chi\partial^{\nu}\chi - \frac{1}{2}\partial_{\nu}\xi\partial^{\nu}\xi -\frac{\alpha^\prime}{16\kappa^2} \mathcal{F}_{\rho\sigma}\mathcal{F}^{\rho\sigma}  - \frac{\alpha^\prime }{8\kappa} \frac{6\sqrt{2}}{4!}\frac{\epsilon^{\alpha\beta\gamma\delta}}{\sqrt{-g}}\xi\mathcal{F}_{\alpha\beta}\mathcal{F}_{\gamma\delta}\bigg{]} \sqrt{-g}~d^{4}x  
\end{align}
where we have considered $e^\chi\simeq 1$ since we are interested in regions where the classical value of the dilaton field is small i.e., $\chi\propto \alpha^\prime$ \cite{Campbell:1992hc}. From \ref{S2-4} it is clear that the axion photon coupling constant is given by $\frac{\alpha^\prime }{8\kappa} \frac{6\sqrt{2}}{4!}$. 
Further, the action in \ref{S2-4} also reveals that the axion and the dilaton fields are massless which is a characteristic feature of the Lagrangian arising in heterotic string theory compactified on a six-dimensional torus \cite{Gross:1986mw,Sen:1992ua,Sen:1994eb}.

The resultant Maxwell's equations coupled to the axion and the dilaton are given by,
\begin{align}
\nabla_{\mu}\mathcal{F}^{\mu\nu}=-\frac{\kappa}{\sqrt{2}}(\partial_\alpha\xi) (^* \mathcal{F})^{\alpha\nu} ,\label{S2-5}
\end{align}
while equations of motion for the dilaton and the axion fields are respectively given by,
\begin{align}
 \nabla_{\mu}\nabla^{\mu}\chi =-  \frac{\sqrt{2}\alpha^\prime}{16\kappa}\mathcal{F}_{\alpha\beta}\mathcal{F}^{\alpha\beta}, \label{S2-7}\\
 \nabla_{\mu}\nabla^{\mu}\xi =\frac{\alpha^\prime}{8\kappa}\frac{6\sqrt{2}}{4!}\mathcal{F}_{\gamma\sigma}(^*\mathcal{F})^{\gamma\sigma} 
\label{S2-8}
\end{align}
where $(^*\mathcal{F})^{\mu\nu}=\frac{\epsilon^{\mu\nu\alpha\beta}}{\sqrt{-g}}\mathcal{F}_{\alpha\beta}$.
The Einstein's equations assume the form,
\begin{align}
\mathcal{G}_{\mu\nu} = \mathcal{T}_{\mu\nu}(\mathcal{F},\chi,\xi) \label{S2-9}
\end{align}
where, the energy-momentum tensor on the right hand side of \ref{S2-9} is given by,
\begin{align}
\label{S2-10}
&\mathcal{T}_{\mu\nu}(\mathcal{F},\chi,\xi) = \frac{-2}{\sqrt{-g}}\frac{\delta\mathcal{S}(\mathcal{F},\chi,\xi)}{\delta g^{\mu\nu}}\nonumber\\
\end{align}
It turns out that the Einstein's equations with the axion, the dilaton and the Maxwell field as the source give rise to an exact, stationary and axisymmetric black hole solution popularly known as the
Kerr-Sen solution \cite{Sen:1992ua} in the literature, which when expressed in Boyer-Lindquist coordinates assume the form \cite{Garcia:1995qz,Ghezelbash:2012qn,Bernard:2016wqo},
\begin{align}
\label{S2-11}
ds^{2} &= - \bigg{(}1 - \frac{2\mathcal{M}r}{\tilde{\Sigma}}\bigg{)}~dt^{2} + \frac{\tilde{\Sigma}}{\Delta}(dr^{2} + \Delta d\theta^{2}) - \frac{4a\mathcal{M}r}{\tilde{\Sigma}}\sin^{2}\theta dt d\phi 
&+ \sin^{2}\theta d\phi^{2}\bigg{[}r(r+r_{2}) + a^{2} + \frac{2\mathcal{M}ra^{2}\sin^{2}\theta}{\tilde{\Sigma}}\bigg{]}
\end{align}
where,
\begin{align}
\label{S2-11a}
\tilde{\Sigma} &= r(r + r_{2}) + a^{2}\cos^{2}\theta \tag {11a}\nonumber\\
\Delta &= r(r + r_{2}) - 2\mathcal{M}r + a^{2} \tag {11b}\nonumber
\end{align}
The ADM mass of the above spacetime is denoted by $M_{ADM}=\mathcal{M}-\frac{r_2}{2}$ while $a$ refers to the rotation parameter of the black hole. The dilaton parameter $r_{2} = \frac{q^{2}}{\mathcal{M}}$ where $q=\sqrt{\frac{\alpha^\prime}{8}}Q$ is proportional to the electric charge $Q$ of the black hole and the square root of the inverse string tension $\alpha^\prime$. In the event $q$ vanishes, \ref{S2-11} reduces to the Kerr metric. The event horizon $r_H$ of the above spacetime is obtained by solving for $\Delta=0$ such that,
\begin{align}
\label{S2-11b}
r_H=\mathcal{M}-\frac{r_2}{2} +\sqrt{\bigg(\mathcal{M}-\frac{r_2}{2}\bigg)^2 - a^2}
\end{align} 
From the form of $r_2$ and \ref{S2-11b} it can be shown that $0\leq r_2 \leq 2$ leads to real, positive event horizons and hence black hole solutions.

We note that the spacetime given by \ref{S2-11} is very similar to the Kerr-Newman solution in \gr\ which differs from \ref{S2-11} due to the absence of the coupling of the axion and the dilaton with the Maxwell field.
The solution of the axion and the dilaton fields are respectively given by \cite{Sen:1992ua,Campbell:1992hc},
\begin{align}
 \xi &= \frac{q^{2}}{\sqrt{2}\kappa G\mathcal{M}}\frac{a\cos\theta}{r^{2} + a^{2}\cos^{2}\theta}  \label{S2-12}\\
e^{2\chi} &= \frac{r^{2} + a^{2}\cos^{2}\theta}{r(r + r_{2}) + a^{2}\cos^{2}\theta} \label{S2-13}
\end{align}
while the solution of the $U(1)$ gauge field assumes the form \cite{Sen:1992ua},
\begin{align}
\label{S2-14}
 A=\frac{2\sqrt{2}qr}{\tilde{\Sigma}}\bigg(-dt +a \mathrm{sin}^2\theta d\phi\bigg) 
\end{align}

We note from \ref{S2-12}$-$\ref{S2-14} that all the above three fields vanish for an asymptotic observer as $r\to \infty$. Since the gravity action in Eintein gravity and EMDA gravity are identical and the additional fields present in EMDA gravity vanish asymptotically, the gravitational waves in both the theories travel with the speed of light. This is in accordance with \cite{Moore:2001bv, Chesler:2017khz}.

It is further evident from \ref{S2-12}$-$\ref{S2-14} that the coupling of the axion and the dialton to the Maxwell field is crucial, as without this, the field strengths associated with both these fields will identically vanish (\ref{S2-12} and \ref{S2-13}).  Therefore, although the Kerr-Sen black hole carries electric charge, it esentially originates from the axion-photon coupling and not the infalling charged particles. 
Moreover, the presence of axionic field renders angular momentum to the black hole (\ref{S2-12}). 
From the solution of the axion and the dilaton fields the non-zero components of $H_{\mu\nu\alpha}$ can also be evaluated \cite{Ganguly:2014pwa}. 
When the rotation parameter in \ref{S2-11} vanishes (i.e., in the absence of the axionic field), the resultant spherically symmetric spacetime represents a black hole labelled by its mass, electric charge and the asymptotic value of the dilaton field \cite{Garfinkle:1990qj,Yazadjiev:1999xq}.   
It is interesting to note that the Kerr-Sen background \ref{S2-11} can also be generated by a Newman-Janis transformation \cite{Newman:1965tw} of the aforesaid spherically symmetric spacetime in pure dilaton coupled gravity \cite{Garfinkle:1990qj,Yazadjiev:1999xq}. 
In what follows we will compute the power associated with astrophysical jets and the radiative efficiencies from the continuum spectrum, in the Kerr-Sen background. This will enable us to understand whether such a gravity theory can be instrumental in explaining these observations. 

\section{Observational avenues to test the Kerr-Sen spacetime}
\label{S3}
In this section we will consider two observational avenues to test the nature of the background spacetime, namely, the continuum spectrum emitted from the accretion disk surrounding the black hole and the power associated with the transient jets observed in such systems. Jets and accretion are ubiquitous to astrophysical systems such as active galactic nuclei and microquasars. \emph{Transient} or \emph{ballistic} jets consist of blobs of radio or X-ray emitting plasma moving ballistically outward with relativistic velocities. They are believed to be launched very close to the event horizon \cite{Mirabel:1999fy} and hence it is expected that the power associated with the transient jets will be affected by the nature of the background metric.

The background spacetime also affects the continuum spectrum from the accretion disk whose peak emission originates very close to the marginally stable circular orbit. The Novikov-Thorne model which is based on the `thin-disk approximation' \cite{Novikov:1973kta} is often used to theoretically mimick the observed sectrum. This approximation holds good primarily when the black hole dwells in the High/Soft state during the outbursts. 

One may also explore superradiance due to scalar fields in the Kerr-Sen background. When we consider superradiant instability of a scalar field in a given background, the scalar field is treated as a perturbation to the given metric. The equation of motion of the scalar field is solved in the said background assuming an ansatz for the scalar field and the energy flux at the horizon is calculated. The flux tends to diverge below a given frequency which depends on the mass of the scalar field.
The onset of superradiance causes the black hole to spin down. Therefore, if the spin of the black hole does not change over a long time scale (say, decades) then it implies that the black hole is stable to superradiance. Comparing with the available observations of black holes in the Regge plane one can therefore establish constrains on the mass of the scalar field \cite{Cardoso:2018tly,Arvanitaki:2010sy,Baryakhtar:2017ngi,Brito:2017wnc,Brito:2017zvb}.

It is important to note that when we consider superradiant instability of a scalar/vector field in a given background, the said field is treated as a perturbation to the given metric. However, the scalar dilaton or the pseudo-scalar axion in \ref{S2-4} are not treated as perturbations to the metric, in fact, they are used as sources to derive the metric and hence these are charges or hairs associated with the black hole. In case one is interested to investigate superradiant instability of scalar/vector bosons in the Kerr-Sen background, then one needs to introduce test fields as perturbation to the Kerr-Sen background. This has been addressed with a test scalar field in \cite{Huang:2017whw} and for a massive vector field in \cite{Cayuso:2019ieu}.

In the next section we will discuss how the continuum spectrum and the power associated with transient jets can be used to probe the background metric. A similar analysis has been performed earlier \cite{Bambi:2012ku} in the context of Johannsen-Psaltis spacetime.

\subsection{Radiative efficiency of black holes from the continuum spectrum} 
\label{S3a}
In this section we highlight the basic features of the Novikov-Thorne model \cite{Novikov:1973kta} which is used to describe the continuum spectrum observed in the black holes. According to this model the electromagnetic emission from the accretion disk surrounding the black hole chiefly contributes to the continuum spectrum. The accretion disk is assumed to be geometrically thin such that matter is accreted chiefly along the equatorial plane. The accreting particles are assumed to maintain nearly circular orbits along the geodesics, with negligible radial velocity arising due to viscous stresses, which facilitates the inspiral and fall of matter into the black hole. 
Since the accreting particles follow nearly circular geodesics the gravitational pull of the central black hole supercedes the forces due to radial pressure gradients. This in turn implies that the specific internal energy of the accreting fluid can be neglected compared to its rest energy such that special relativistic corrections to the local hydrodynamic, thermodynamic and radiative properties of the fluid can be safely ignored compared to the general relativistic effects due to the presence of the black hole.

As the matter falls towards the black hole they lose gravitational potential energy which gets converted into electromagnetic radiation interacting very effectively with the accreting matter before being radiated out of the system. Consequently, the geometrically thin accretion disk is also optically thick and practically no heat is trapped with the accretion flow. Due to the efficient interaction between matter and radiation, every annulus of the disk emits a black body commensurate with the temperature of the disk. The total emission from the accretion disk is therefore a multi-color black body spectrum peaking in soft X-rays for stellar mass black holes. For a more detailed description of the thin-disk model one is referred to \cite{Novikov:1973kta,Page:1974he,Banerjee:2019sae}.
This model provides an accurate description of the observed continuum spectrum when the black hole is in the High/Soft state during the outbursts. In such a scenario the peak emission from the accretion disk generally emerges from the  marginally stable circular orbit. 
The peak temperature and flux of this continuum spectrum are used to estimate the radius of the innermost stable circular orbit $r_{\rm isco}$ of a black hole, provided its mass, distance and inclination angle are known from independent measurements. The radius of the innermost stable circular orbit in turn depends on the background metric and is obtained from the effective potential $V_{\rm eff}$ in which the accreting particles move. The effective potential in a stationary and axi-symmetric spacetime is given by \cite{Novikov:1973kta,Page:1974he,Banerjee:2019sae},
\begin{align}\label{S3-1}
V_{\rm eff}(r)=\frac{E^2g_{\phi\phi}+2ELg_{t\phi}+L^2g_{tt}}{g_{t\phi}^2-g_{tt}g_{\phi\phi}}-1
\end{align}
where, $g_{tt}$, $g_{t\phi}$ and $g_{\phi\phi}$ are the metric elements given in \ref{S2-11} while $E$ and $L$ are the specific energy and specific angular momentum of the particles such that,
\begin{align}\label{S3-2}
E=\frac{-g_{tt}-\Omega g_{t\phi}}{\sqrt{-g_{tt}-2\Omega g_{t\phi}-\Omega^2 g_{\phi\phi}}}~.
\end{align}
and 
\begin{align}\label{S3-3}
L=\frac{\Omega g_{\phi\phi}+g_{t\phi}}{\sqrt{-g_{tt}-2\Omega g_{t\phi}-\Omega^2 g_{\phi\phi}}}~,
\end{align}
where the angular velocity $\Omega=(d\phi/dt)$ of the test particles is given by,
\begin{align}\label{S3-4}
\Omega=\frac{d\phi}{dt}=\frac{-g_{t\phi,r}\pm \sqrt{\left\lbrace-g_{t\phi,r}\right\rbrace^2-\left\lbrace g_{\phi\phi,r}\right\rbrace \left\lbrace g_{tt,r}\right\rbrace}}{g_{\phi\phi,r}}~.
\end{align} 

The radius of the innermost stable circular orbit corresponds to the inflection point of this effective potential such that $V_{\rm eff}=\partial _{r}V_{\rm eff}=0=\partial _{r}^{2}V_{\rm eff}$ \cite{Banerjee:2019sae}.
Therefore, a measurement of $r_{isco}$ from the continuum spectrum can be used to constrain the background spacetime. In particlular, if the background is taken to be Kerr spacetime, then a measurement of $r_{isco}$ from the continuum spectrum can be used to predict the angular momentum of the black holes \cite{Brenneman:2013oba}. This forms the basis of the Continuum Fitting Method used to determine the black hole spins \cite{McClintock:2013vwa}.

\begin{figure}[t!]
\begin{center}
\subfloat[\label{fig1a}]{\includegraphics[scale=0.6]{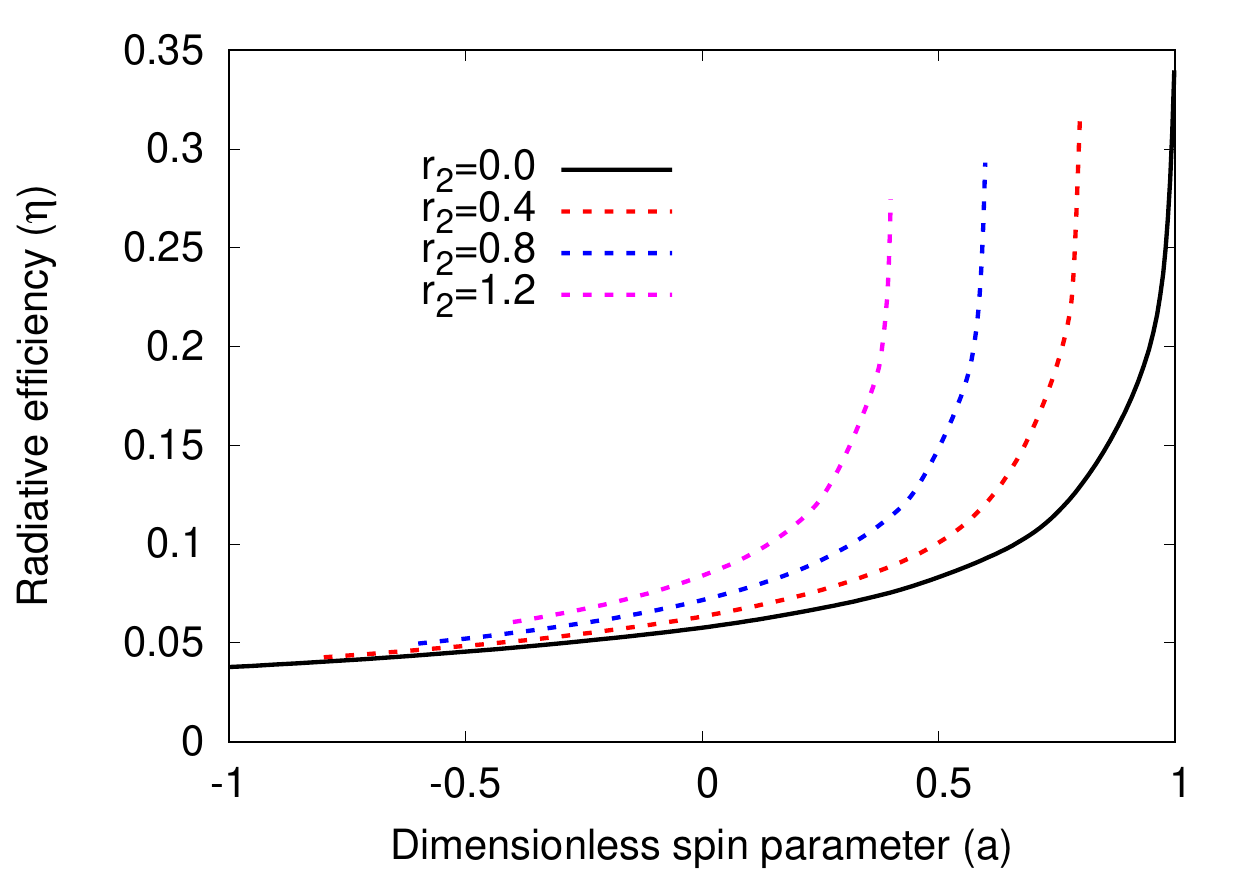}}
\hspace{0.5cm}
\subfloat[\label{fig1b}]{\includegraphics[scale=0.6]{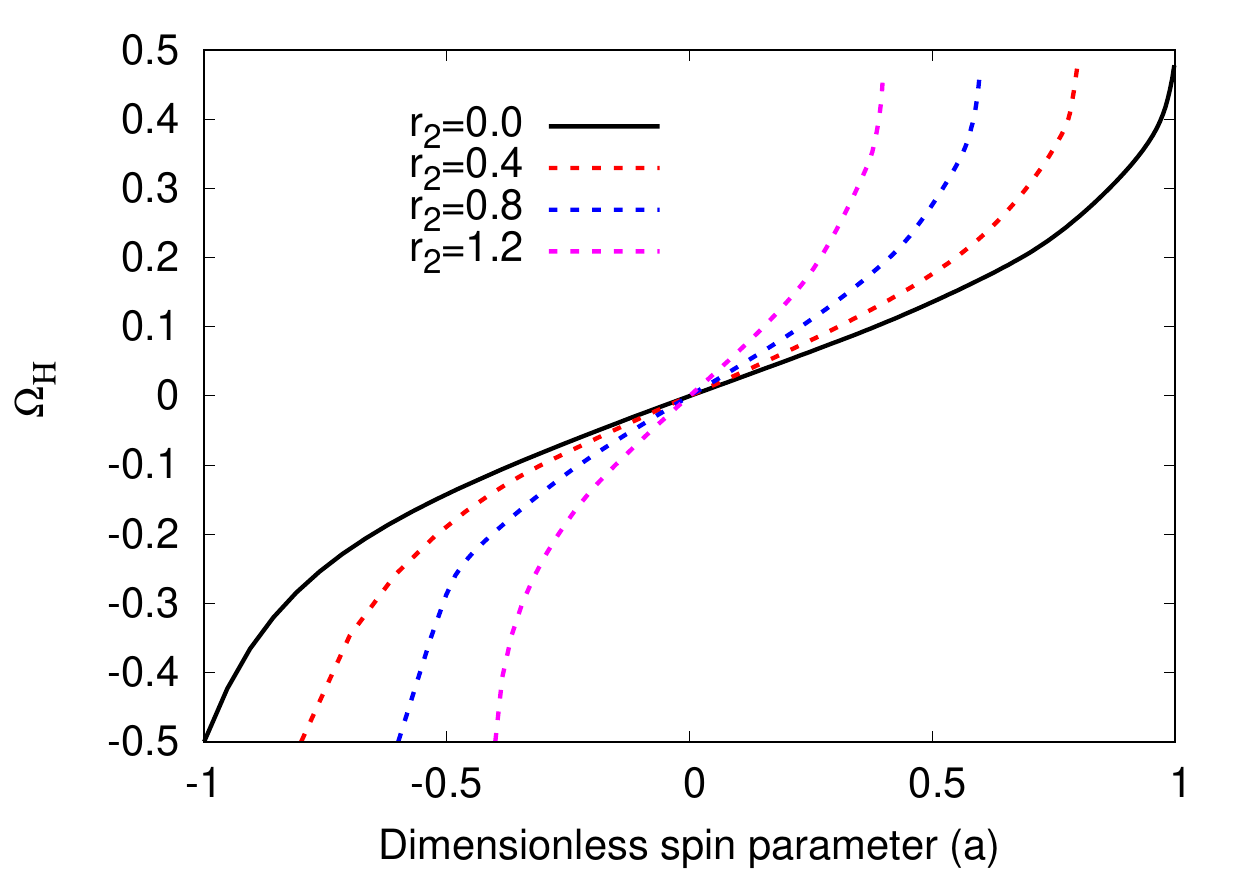}}
\caption{The above figure illustrates the variation of (a) the radiative efficiency $\eta$ and (b) the angular velocity of the horizon $\Omega_H$ with the dimensionless spin parameter $a$ for various choices of the dilaton parameter $r_2$. The black solid line corresponds to the Kerr scenario.}
\label{Fig1}
\end{center}
\end{figure}

The Continuum Fitting Method eventually determines the radiative efficiency $\eta$ of a black hole which corresponds to the gravitational binding energy of a test particle at the innermost stable circular orbit, such that 
\begin{align}
\label{S3-5}
\eta=1-E_{\rm isco}
\end{align}
where $E_{\rm isco}$ is the specific energy of the test particle computed at $r_{isco}$. It is evident from \ref{S3-5} that $\eta$ also depends on the background metric and if 
the spin of the black hole is determined by the Continuum Fitting Method then $\eta$ can also be evaluated.  \\
In the event we consider departure from \gr, the radiative efficiency computed from the continuum emission by the above method can be used to determine the allowed values of the metric parameters for a given black hole.  \ref{fig1a} depicts the variation of the radiative efficiency $\eta$ (\ref{S3-5}) with the dimensionless spin parameter $a$ for various values of the dilaton parameter $r_2$ (Here and in the rest of the paper the spin and the dilaton parameter are scaled by the mass of the black hole, i.e. $r_2\equiv r_2/\mathcal{M}$ and $a\equiv a/\mathcal{M}$.) We note that for a given $r_2$, $\eta$ increases with $a$.

\subsection{Jets in astrophysical systems and the Blandford-Znajeck model}
\label{S3b}
Microquasars generally exhibit two different types of jets \cite{Fender:2004aw}: (a) \emph{Steady}, non-relativistic jets (also known as outflows) which are common during the hard-state \cite{Markoff:2005ht} and are observed at a wide range of accretion luminosities and (b) \emph{Transient} or \emph{ballistic} jets which occur at the Eddington luminosity during state transitions, mainly when the source transits from the hard to the soft state at high accretion rates. Transient jets which reach out to parsec scales 
are relativistic in nature and are believed to be launched very close to the event horizon \cite{Mirabel:1999fy}. Hence these jets are often instrumental in extracting large fractions of the spin energy of the black holes \cite{Narayan:2011eb}.  
Since the main goal of this work is to constrain the Kerr-Sen metric from the jet power, we will be concentrating on the relativistic transient jets in this work.  

The exact mechanism of generating the transient jets is unknown, although a number of theoretical models \cite{1990ApJ...354..583P,1990ApJ...363..206T,Koide:2003fj} have been proposed that can potentially explain the observed jet power. 
One of the popular models used for explaining the formation of relativistic jets was put forward by Blandford and Znajeck \cite{Blandford:1977ds} where the relativistic jets are powered by extracting the rotational energy of the black holes by means of magnetic field lines which are supported by the surrounding accretion disk due to the presence of electric currents. 
The Blandford-Znajeck model was originally proposed for Kerr black holes surrounded by a stationary, axi-symmetric force-free magnetosphere. However, this can be generalized to any stationary, axi-symmetric spacetime.

The force-free magnetosphere has the property that the particle inertia is negligible such that the total energy momentum tensor is dominated by the energy-momentum tensor due to the electromagnetic fields, i.e., 
\begin{align}
\label{1}
T_{\mu\nu}^{tot}\approx T_{\mu\nu}^{EM}=F_{\mu\rho} F^\rho_\nu- \frac{1}{4} g_{\mu\nu}F^{\alpha\beta}F_{\alpha\beta}
\end{align}  
which satisfies the conservation equation,
\begin{align}
\label{2}
\nabla^\mu T_{\mu\nu}^{EM}=0
\end{align}
In \ref{2}, 
$F_{\mu\nu}=\partial_\mu A_\nu-\partial_\nu A_\mu$ is the Faraday tensor and $A_\mu$ is the gauge field. 
In a force-free magnetosphere it can be shown that,
\begin{align}
\label{3}
\frac{A_{t,r}}{A_{\phi,r}}=\frac{A_{t,\theta}}{A_{\phi,\theta}}=-\omega(r,\theta)
\end{align}
where $\omega(r,\theta)$ represents the electromagnetic angular velocity \cite{Blandford:1977ds}. With this  force-free condition (\ref{3}) and assuming $A_\mu$ is axi-symmetric and time independent, one can write the Faraday tensor in the form,
\begin{equation}
\label{4}
F_{\mu\nu}=\sqrt{-g}
\begin{pmatrix}
0 & -\omega B^\theta & \omega B^r & 0 \\
\omega B^\theta & 0 & B^\phi & -B^\theta \\
-\omega B^r & -B^\phi & 0 & B^r \\
0 & B^\theta & -B^r & 0 
\end{pmatrix}
\end{equation}

It can be shown that the power associated with the relativistic jet in the context of the Blandford-Znajeck model is given by (\ref{AA} and \ref{AB}),
\begin{align}
\label{4}
P_{BZ}=4\pi \int_0^{\pi/2} \sqrt{-g}~T^r_t~d\theta
\end{align}
where, $T^r_t$ represents the radial component of the Poynting flux evaluated at the jet lauching radius which happens to be the event horizon. This is given by,
\begin{align}
\label{5}
T^r_t=2r_H M sin^2\theta(B^r)^2\omega \big[\Omega_H-\omega\big]\biggr\rvert_{r=r_{H}}
\end{align}
where, $r_H$ and $\Omega_H=a/(2Mr_H)$ are the horizon radius and the angular velocity of the event horizon, respectively. 

At this stage, it is impossible to calculate the power $P_{BZ}$ associated with the jet without knowing the form of $\omega$ and $B^r$. Ideally this should be obtained by solving \ref{2}, which is quite non-trivial. 
Therefore, we follow the standard approach \cite{Tanabe:2008wm,Tchekhovskoy:2009ba}, where an exact solution of \ref{2} is obtained for the Schwarzschild spacetime and then an expansion in $\Omega_H$ is considered to find the rotating solution perturbatively. With this expansion, the jet power in the Blandford-Znajeck model at the leading order in $\Omega_H$ is given by,
\begin{align}
\label{6}
P_{BZ}=k\Phi_{tot}^2 \Omega_H^2
\end{align}
where, $k=1/6\pi$ for a split monopole field profile and $k=0.044$ for a paraboloidal profile. In \ref{6}, $\Phi_{tot}$ denotes the magnetic flux threading the event horizon and is given by,
\begin{align}
\label{7}
\Phi_{tot}=2\pi\int_0^{\pi}\sqrt{-g}|B^r| d\theta
\end{align} 
For a more detailed derivation of the jet power in the Blandford-Znajeck model assuming Kerr-Sen background, one is referred to \ref{AA} and \ref{AB}. We note that the dependence of $P_{BZ}$ on the metric arises through $\Omega_H$. In \ref{fig1b} we plot the variation of $\Omega_H$ with the dimensionless spin parameter $a$ for various values of $r_2$. The figure shows that for a given $r_2$, $|\Omega_H|$ increases with $|a|$.

\section{Comparison of the theoretical model with observations }
\label{S4}

We have noted in \ref{S3} that the radiative efficiency (\ref{S3-5}) and the jet power (\ref{6}) are both sensitive to the background metric. Therefore, if these quantities are observationally constrained for some of the black holes, it can be used to gain some insight on the observationally favored magnitude of the dilaton parameter $r_2$.  
Our observational sample comprises of 
six X-ray binaries, namely, GRS1915+105, GROJ1655-40, XTEJ1550-564, A0620-00, H1743-322 and GRS1124-683 whose jet power and radiative efficiency are known from observations \cite{Pei:2016kka,Narayan:2011eb,Steiner:2012ap}. 

The spins of these microquasars have been estimated by the Continuum Fitting method which in turn have been used to evaluate the radiative efficiencies of these black hole sources. The mass $M$, the distance $D$, the inclination angle $i$, the dimensionless spin (Kerr parameter) $a$ and the radiative efficiency $\eta$ of these black holes are reported in \ref{Table1} \cite{Pei:2016kka}.

For the six microquasars, we follow the prescription of \cite{Narayan:2011eb,Steiner:2012ap} to determine the observed jet power which assumes that the entire power in the transient jet is proportional to the peak 5 GHz radio flux density $(S_{\nu,0})_{ max,~ 5GHz}$, also reported in \ref{Table1}. 
This observed flux density needs to be appropriately Doppler boosted for both the approaching and the receding jets and summed to obtain the corresponding emitted flux density \cite{Steiner:2012ap,Mirabel:1999fy}. This is scaled by the distance of the black hole to obtain the luminosity and by the black hole mass to remove any dependence. Using the natural units for these systems the proxy for the jet power is given by \cite{Narayan:2011eb,Steiner:2012ap},
\begin{align}
\label{8}
P_{jet}=\bigg(\frac{\nu}{\mathrm{5~GHz}}\bigg)\bigg(\frac{S_{\nu,0}^{tot}}{\mathrm{Jy}}\bigg)\bigg(\frac{D}{\mathrm{kpc}}\bigg)^2\bigg(\frac{M}{M_\odot}\bigg)^{-1}
\end{align}
where, $\nu S_{\nu,0}^{tot}$ is the beaming corrected maximum flux after taking into account the approaching and receding jets \cite{Steiner:2012ap,Mirabel:1999fy}. In order to correct for the beaming the Lorentz factor $\Gamma$ associated with the jet is taken to be $2\lesssim \Gamma\lesssim 5$ \cite{Fender:2004gg,2006csxs.book..381F}, commensurate with the mildly relativistic jets in microquasars.

\begin{table}[t!]
\vskip0.2cm
{\centerline{\large Table 1}}
{\centerline{}}
\caption{Parameters of the transient black hole binaries}
\label{Table1}
{\centerline{}}
\begin{center}
\begin{tabular}{|c|c|c|c|c|c|c|}

\hline
$\rm BH~ Source$ & $ M (M_\odot)$ & $ D (kpc)$ & $ i^\circ$ & $ a$ & $\eta$ &$\rm (S_{\nu,0})_{ max,~ 5GHz} (Jy)$  \\
\hline 
$\rm A0620-00$ & $\rm 6.61\pm 0.25$ & $\rm 1.06\pm 0.12$ & $\rm 51.0\pm 0.9$ & $\rm 0.12\pm 0.19$ & $\rm 0.061^{+0.009}_{-0.007}$ & $\rm 0.203$ \\ \hline
$\rm H1743-322$ & $\rm 8.0$ & $\rm 8.5\pm 0.8$ & $\rm 75.0 \pm 3.0$ & $\rm 0.2\pm 0.3$ & $\rm 0.065^{+0.017}_{-0.011}$ & $\rm 0.0346$    \\ \hline
$\rm XTE J1550-564$ & $\rm 9.10\pm 0.61$ & $\rm 4.38\pm 0.5$ & $\rm 74.7\pm 3.8$ & $0.34\pm 0.24$ & $0.072^{+0.017}_{-0.011}$ & $\rm 0.265$  \\ \hline
$\rm GRS 1124-683$ & $\rm 11.0^{+2.1}_{-1.4}$ & $\rm 4.95^{+0.69}_{-0.65}$ & $\rm 43.2^{+2.1}_{-2.7}$ & $\rm 0.63^{+0.16}_{-0.19}$ & $\rm 0.095^{+0.025}_{-0.017}$ & $\rm 0.45$  \\ \hline
$\rm GRO J1655-40$ & $\rm 6.30\pm 0.27$ & $\rm 3.2\pm 0.5$ & $\rm 70.2\pm 1.9$ & $\rm 0.7\pm 0.1$ & $\rm 0.104^{+0.018}_{-0.013}$ & $\rm 2.42$ \\ \hline
$\rm GRS 1915+105$ & $\rm 12.4^{+1.7}_{-1.9}$ & $\rm 8.6^{+2.0}_{-1.6}$ & $\rm 60.0\pm 5.0$ & $\rm 0.975~ a_*>0.95$ & $\rm 0.224~ \eta>0.19$ & $\rm 0.912$  \\ \hline
\end{tabular}
\end{center}
\end{table}

\begin{table}[t!]
\vskip0.2cm
{\centerline{\large Table 2}}
{\centerline{}}
\caption{Proxy jet power values in units of $\rm kpc^2GHz Jy M_\odot^{-1}$}
\label{Table2}
{\centerline{}}
\begin{center}
\begin{tabular}{|c|c|c|}

\hline
$\rm BH~ Source$ & $\rm  P_{jet}\rvert_{\Gamma=2} $ & $\rm P_{jet}\rvert_{\Gamma=5}$  \\
\hline 
$\rm A0620-00$ & $\rm 0.13$ & $\rm 1.6$  \\ \hline
$\rm H1743-322$ & $\rm 7.0$ & $\rm 140$  \\ \hline
$\rm XTE J1550-564$ & $\rm 11$ & $\rm 180$  \\ \hline
$\rm GRS 1124-683$ & $\rm 3.9$ & $\rm 390$  \\ \hline
$\rm GRO J1655-40$ & $\rm 70$ & $\rm 1600$  \\ \hline
$\rm GRS 1915+105$ & $\rm 42$ & $\rm 660$  \\ \hline

\end{tabular}

\end{center}
\end{table}

Assuming the Lorentz factors of $\Gamma=2$ and $\Gamma=5$ and using \ref{8}, the Doppler corrected jet powers for the six black hole sources are reported in \ref{Table2}\cite{Pei:2016kka,Middleton:2014cha} which are used for comparison with the theoretically derived jet power given by \ref{6}. We note that in \ref{6}, the dependence of the jet power on the metric comes through the term $\Omega_H^2$, while the remaining terms depend on the nature and properties of the magnetic field threading the event horizon. We rewrite \ref{6} in the form,
\begin{align}
\textrm{log} P=\textrm{log} K +2 \textrm{log} \Omega_H\rm , 
\label{9}
\end{align}
where the magnitude of $K$ has been estimated \cite{Middleton:2014cha,Narayan:2011eb} by fitting \ref{9} to the observed jet power plotted against $\Omega_H$, which in turn is calculated from the spin estimated by the Continuum Fitting Method (\ref{S3b}). Since the jet power depends on the Lorentz factor $\Gamma$, the magnitude of $K$ varies accordingly. It turns out that for $\Gamma=2$ and $\Gamma=5$, log$ K=2.94\pm 0.22$ and log$K=4.19\pm 0.22$ respectively, at $90\%$ confidence level \cite{Middleton:2014cha}. In what follows we continue to use these values of $K$ while constraing the metric parameters $r_2$ and $a$ from the observed jet power, as $K$ is independent of the background spacetime.

\subsection{Results}
\begin{itemize}
\item {\bf A0620-00:} The X-ray binary A0620-00 comprises of a K-Type main sequence star and a black hole of $6.6 M_\odot$ \cite{Cantrell:2010vh}. It is the nearest known X-ray binary to the solar system \cite{Foellmi:2008gr} and has an orbital period of $7.75$ hours\cite{1986ApJ...308..110M,Gou:2010qq}. The distance and inclination of the source are reported in \ref{Table1} \cite{Cantrell:2010vh}. The spin of the source has been determined by the Continuum-Fitting method with $-0.59<a<0.49$, the best-fitting value being $a=0.12\pm 0.19$ \cite{Gou:2010qq} which in turn enables us to compute its radiative efficiency $\eta$ (\ref{Table1}). The blue shaded region in \ref{F1} represents the allowed values of $r_2$ and $a$ which can explain the radiative efficiency of this source, within the error bars. The blue solid line corresponds to the contour in the $r_2-a$ plane when the theoretical radiative efficiency given by \ref{S3-5} coincides with the central value of the observed $\eta$ (\ref{Table1}). The blue dotted lines are    
similarly associated with the error bars in the observed $\eta$.

\begin{figure}[t!]
\begin{center}
\subfloat[\label{f1a}]{\includegraphics[scale=0.45]{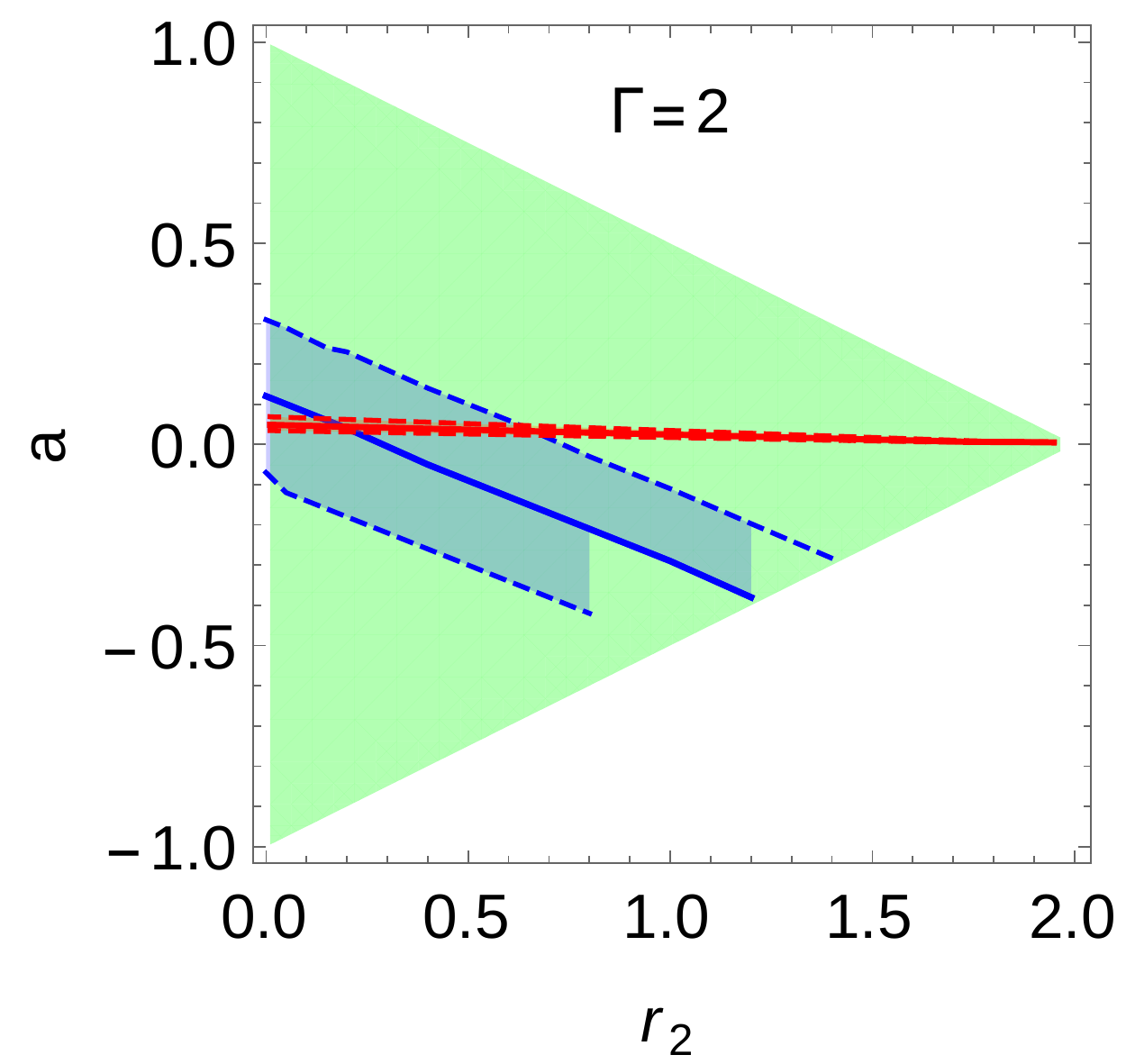}}
\hspace{0.5cm}
\subfloat[\label{f1b}]{\includegraphics[scale=0.45]{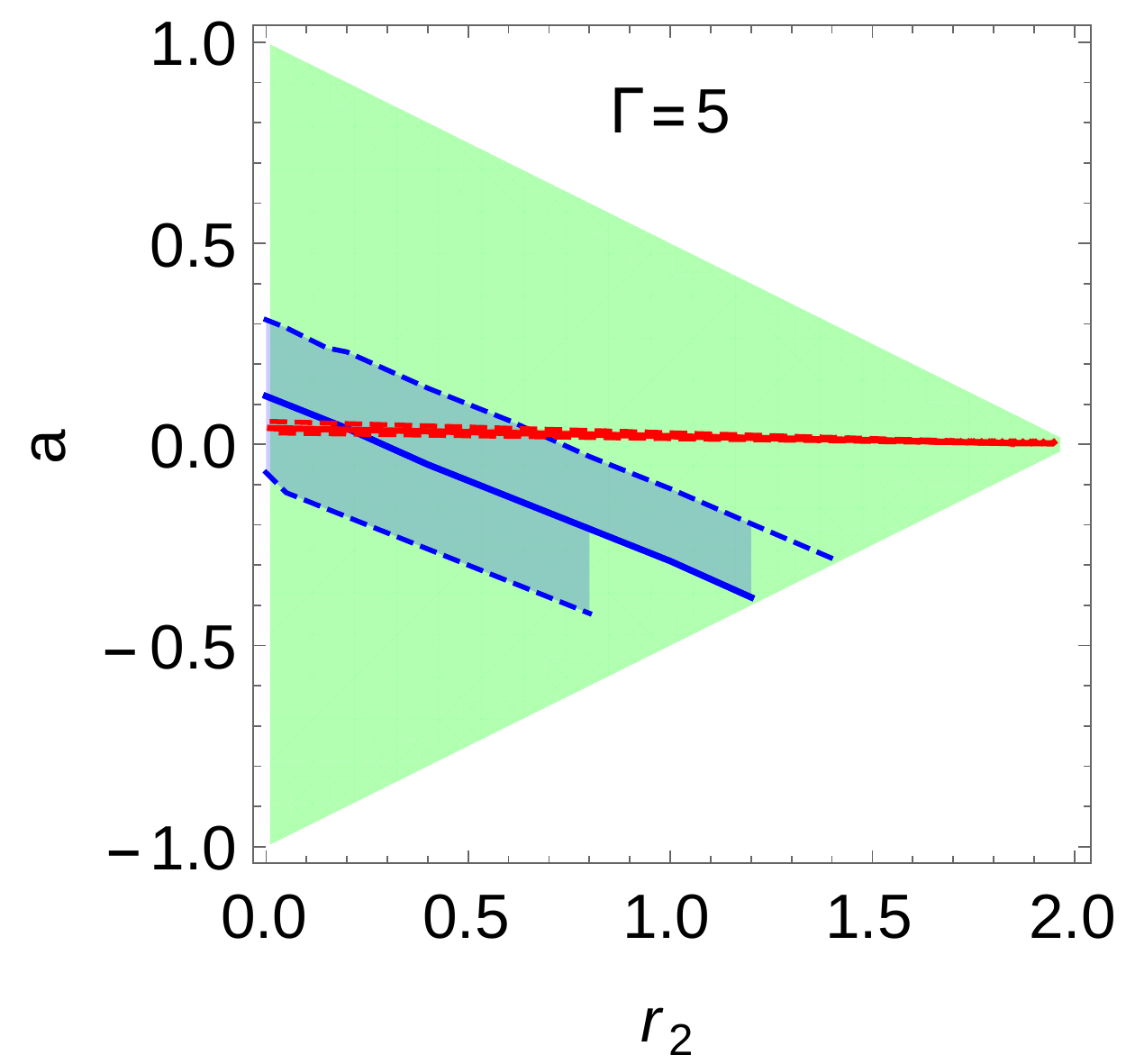}}
\caption{Black hole source A0620-00: The orange shaded area denotes the values of $r_2$ and $a$ when the theoretical jet power $P_{BZ}$ equals the observed luminosity Doppler boosted by the Lorentz factor with (a) $\Gamma=2$  and (b) $\Gamma=5$. The solid red line corresponds to the contour in the $r_2-a$ plane when $P_{BZ}$ equals $P_{jet}$ (\ref{Table2}), while the dashed red lines denote the error bar of $0.3 ~dex$ about $P_{jet}$. 
\newline
The blue shaded region in the figure denote the values of $r_2$ and $a$ when the observed $\eta$ is reproduced by the theoretically calculated radiative efficiency. The blue solid line corresponds to the central value of the observed $\eta$ while the blue dashed lines represent the associated error bars (\ref{Table1}). 
The green shaded region indicates the values of $r_2$ and $a$ giving rise to a real positive event horizon and hence a black hole solution. For more discussion see text.}
\label{F1}
\end{center}
\end{figure}

 
Radio observations of the object reveal the presence of strong radio jets \cite{Kuulkers:1999kn,Gou:2010qq}, the $5~\rm GHz$ radio flux density being $0.203 \rm Jy$ \cite{Narayan:2011eb} (\ref{Table1}). As discussed earlier, the observed radio flux density is converted to the emitted radio luminosity by Doppler deboosting with Lorentz factors $\Gamma=2$ and $\Gamma=5$, the putative values being reported in \ref{Table2}. These are then compared with the theoretical jet power $P_{BZ}$ (given by \ref{6}) to discern the allowed values of $r_2$ and $a$ from jet related observations. An error of $0.3  ~dex$ is considered in the observed jet power $P_{jet}$ \cite{Narayan:2011eb,Middleton:2014cha}. The orange shaded region in \ref{F1} depicts the allowed values of $r_2$ and $a$ that can explain the observed jet power within the error bars. Again the solid red line depicts the contour in the $r_2-a$ plane which can reproduce the central value of $P_{jet}$ while the dashed red lines represent the values of $r_2$ and $a$ that can explain the jet power with error of $0.3  ~dex$ about the central value.

The results for $\Gamma=2$ and $\Gamma=5$ are depicted in \ref{f1a} and \ref{f1b} respectively. The green shaded region denotes the parameter space in the $r_2-a$ plane with real positive event horizons which leads to black hole solutions in EMDA gravity. 
In the subsequent discussion, the definition of the blue, orange and the green shaded region remains the same for the remaining X-ray binaries. 
From \ref{F1} we note that the observed $\eta$ cannot be explained if $r_2>1.5$. The jet power on the other hand can be reproduced by almost the entire range of $r_2$ although the the Kerr parameter varies between: $0\lesssim a \lesssim 0.05$. The intersection of the blue and the orange shaded region represents the allowed values of $r_2$ and $a$ such that both the observations related to $P_{jet}$ and $\eta$ can be explained. From \ref{F1} we note that $0\lesssim r_2\lesssim 0.8$ can describe both the aforesaid observations. Moreover, the allowed ranges of spin from both the observations exhibit an overlap in the general relativistic scenario ($r_2=0$). \\

\item {\bf H1743-322:} This galactic microquasar is located at a distance of $8.5\pm 0.8$ kpc and has an inclination of $75\pm 3^\circ$ \cite{Steiner:2011kd}. Although the mass of this object has not been dynamically measured it has been  predicted to be in the range $8-13 M_\odot$ \cite{Petri:2008jc,Pei:2016kka}. The companion star consists of a late-type main sequence star located in the galactic bulge \cite{Chaty:2015pda} and the orbital period of the binary is 10 hours \cite{Jonker:2009sk}. The spin of the object estimated by the Continuum-Fitting method turns out to be $0.2\pm 0.3$ at $68\%$ confidence and $-0.3<a<0.7$ at $90\%$ confidence \cite{Steiner:2011kd}. The corresponding radiative efficiency is reported in \ref{Table1}. As before, the blue shaded region in \ref{F2} is associated with the allowed values of $r_2$ and $a$ that can describe the observed radiative efficiency within the error bars. The blue lines denote the contours in the $r_2-a$ plane when the observed eta is reproduced by the theoretical radiative efficiency given by \ref{S3-5} (solid blue line for the central value and the dashed blue lines describe the errors about the central value \ref{Table1}). From \ref{F2} it is evident that $r_2>1.6$ cannot explain the observed $\eta$.  
 
\begin{figure}[t!]
\begin{center}
\subfloat[\label{f2a}]{\includegraphics[scale=0.45]{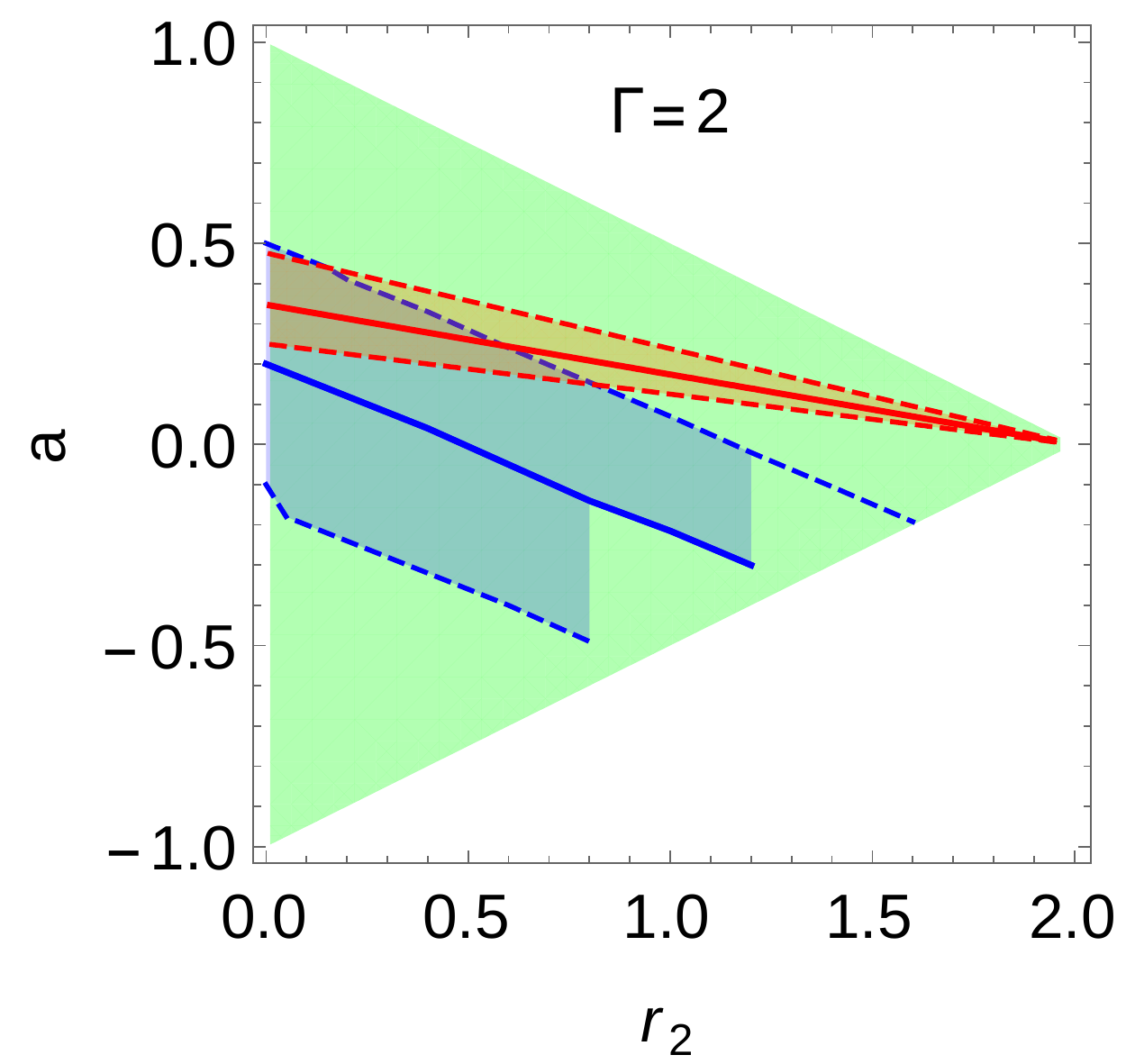}}
\hspace{0.5cm}
\subfloat[\label{f2b}]{\includegraphics[scale=0.45]{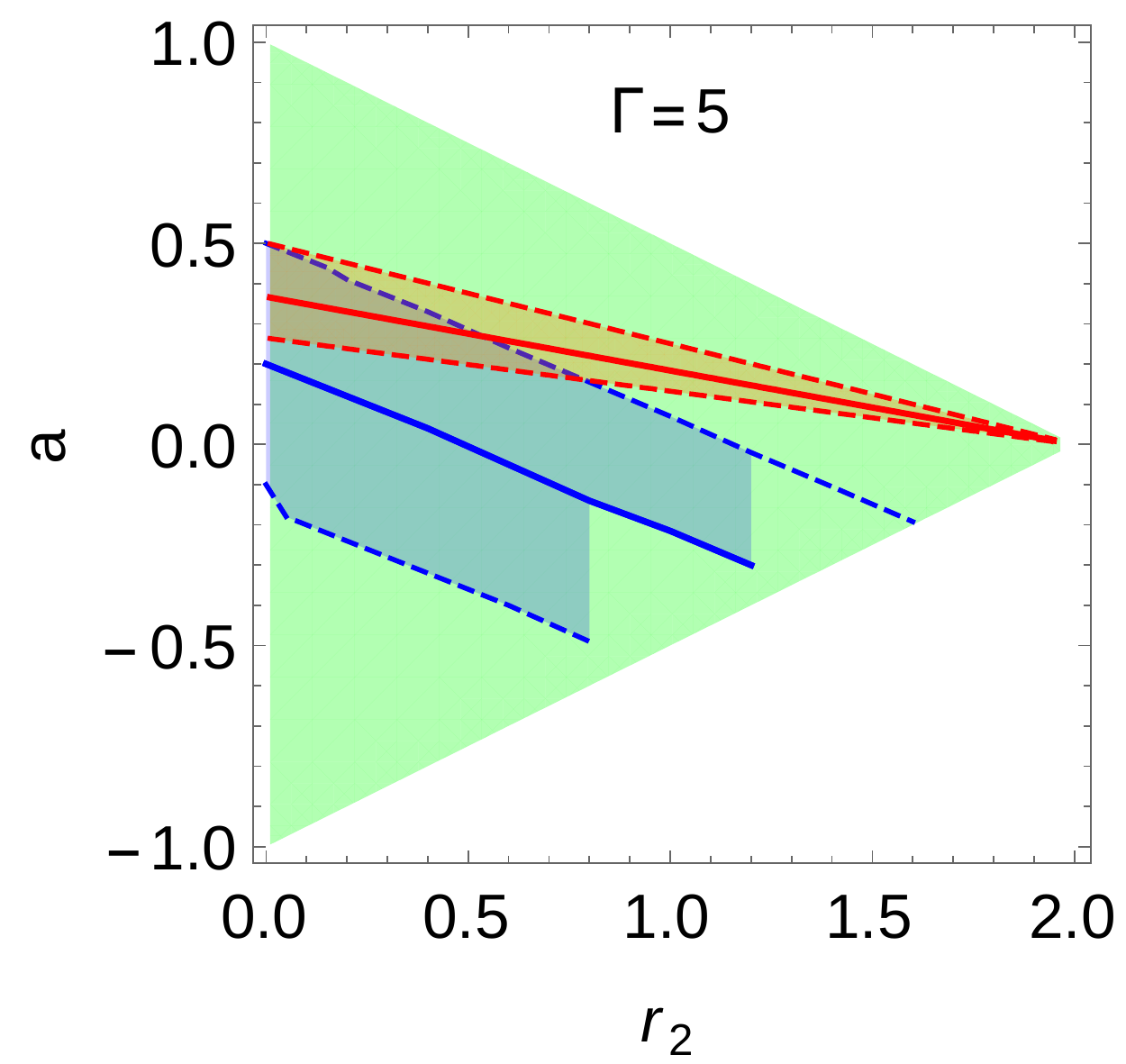}}
\caption{Black hole source H1743-322: The orange shaded region in the figure denote the values of $r_2$ and $a$ that can reproduce the observed $P_{jet}$ with (a) $\Gamma=2$  and (b) $\Gamma=5$, within the error bars.
\newline
The blue shaded region is associated with the allowed values of $r_2$ and $a$ that can address the observed $\eta$ (\ref{Table1}) within the error bars. For more discussion see text.
}
\label{F2}
\end{center}
\end{figure}
The object exhibits strong ballistic jets \cite{Steiner:2011kd} and the emitted jet power corresponding to $\Gamma=2$ and $\Gamma=5$ are reported in \ref{Table2}. These are associated with an error of  $0.3  ~dex$ about the central value \cite{Narayan:2011eb, Middleton:2014cha}. 
In \ref{F2} the allowed values of $r_2$ and $a$ that can explain the emitted jet power along with the positive and the negative errors are depicted by the orange shaded region. The definition of the red solid and dashed lines remain identical to the previous case. The emitted jet power corresponding to $\Gamma=2$ and $\Gamma=5$ are reported in \ref{f2a} and \ref{f2b} respectively.
We note from \ref{F2} that the allowed values of spin from the observed jet power and the radiative efficiency, exhibit an overlap in the general relativistic scenario ($r_2=0$). We further note that almost the entire allowed range of $r_2$ can describe the emitted jet power and the restriction on $r_2$ actually arises from the observed $\eta$. Again the zone of intersection between the blue and the orange shaded region represents the values of $r_2$ and $a$ that describes both the observations. From \ref{f2a} and \ref{f2b} we note that the allowed values of $r_2$ correspond to $0\lesssim r_2\lesssim 0.8$, which interestingly coincides with the range allowed by the previous source. \\

\item {\bf XTE J1550-564:} XTE J1550-564 consists of a binary system with a black hole of mass $9.1\pm 0.61 M_\odot$ \cite{Orosz:2011ki} and a late G or early K-type star as the companion \cite{Orosz:2001qd}. The orbital period of the binary is 1.55 days \cite{Orosz:2001qd}. The distance and inclination of the source are $4.38^{+0.58}_{-0.41}$ kpc and $74.7\pm 3.8^\circ$ respectively \cite{Orosz:2011ki}. The spin of the black hole has been estimated both by the Continuum Fitting and the Fe-line method. The result obtained from the Continuum Fitting method corresponds to $-0.11<a<0.71$ ($90\%$ confidence)\cite{Steiner:2010bt}, with a most likely spin of $a=0.34$ while Fe-line method gives a spin estimate of $a=0.55^{+0.15}_{-0.22}$ \cite{Steiner:2010bt}. In \ref{Table1} the spin corresponding to the Continuum Fitting method has been reported and $\eta$ is calculated based on this result \cite{Pei:2016kka}. 
\begin{figure}[t!]
\begin{center}
\subfloat[\label{f3a}]{\includegraphics[scale=0.45]{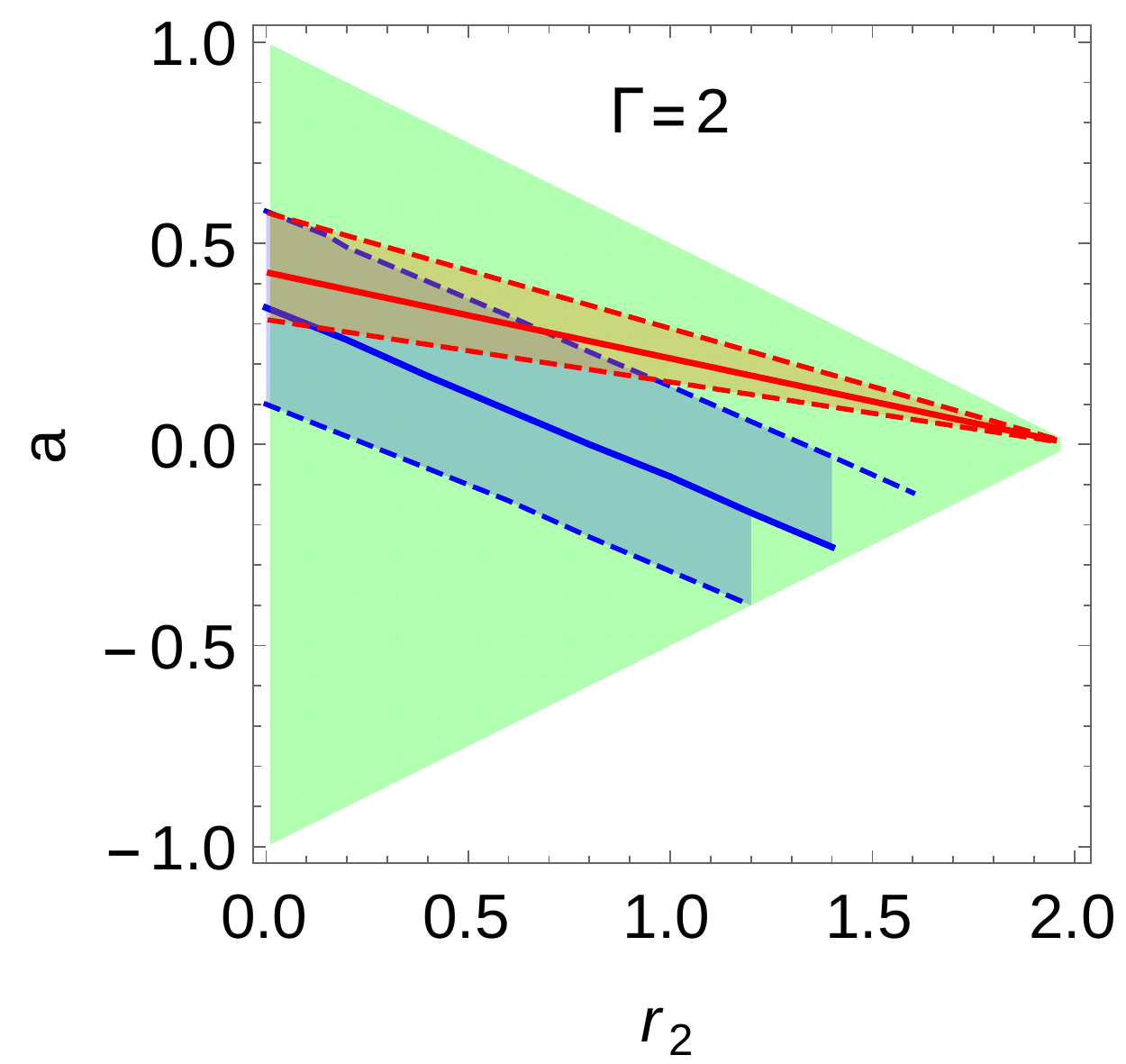}}
\hspace{0.5cm}
\subfloat[\label{f3b}]{\includegraphics[scale=0.45]{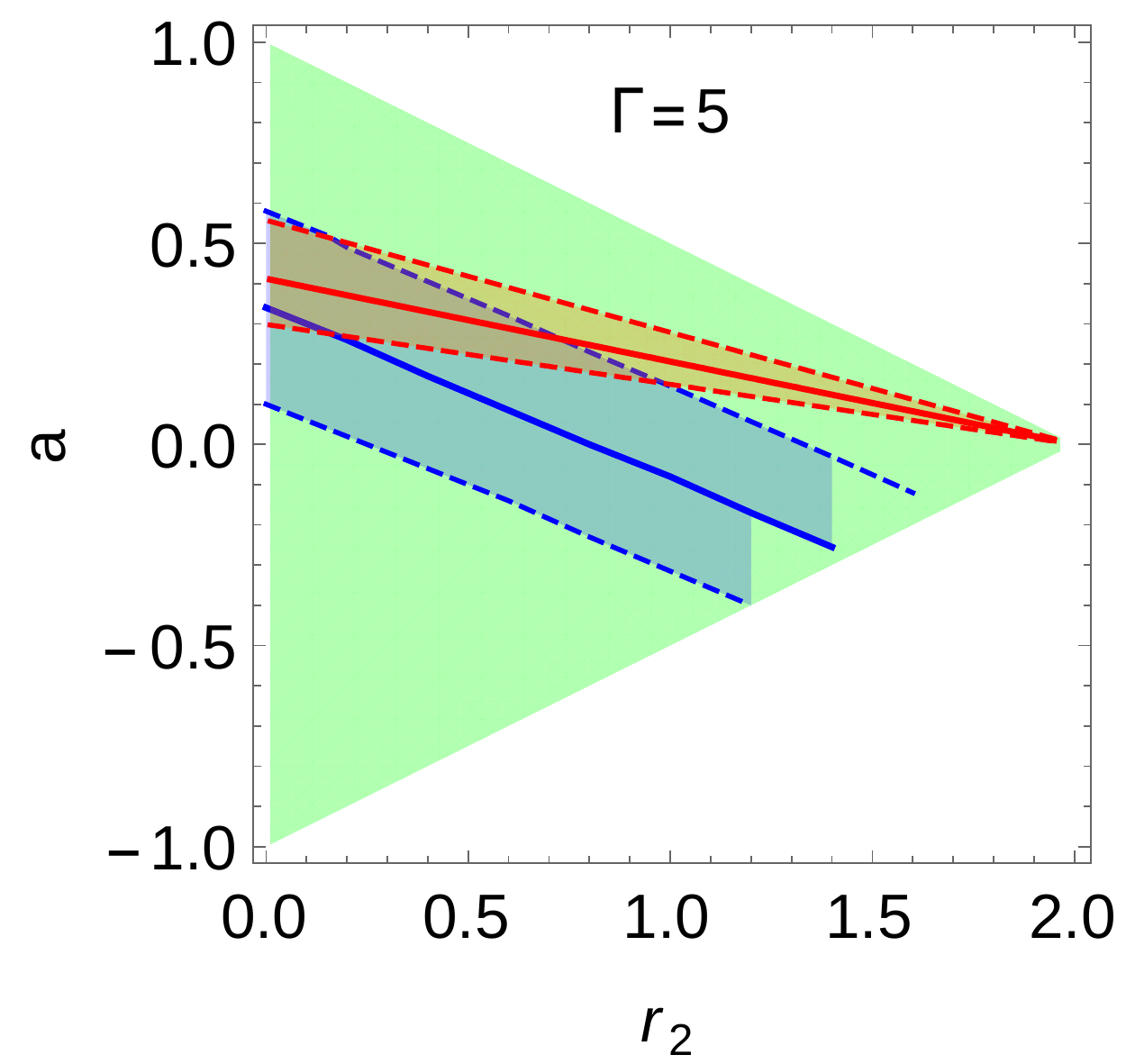}}
\caption{Black hole source XTE J1550-564: The orange shaded region represents the values of $r_2$ and $a$ when the theoretical jet power explains the observed luminosity Doppler boosted by the Lorentz factor (a) $\Gamma=2$ and (b) $\Gamma=5$. 
The blue shaded region is associated with the allowed values of $r_2$ and $a$ that can address the observed $\eta$ (\ref{Table1}) within the error bars.
The green shaded region indicates the values of $r_2$ and $a$ giving rise to a real positive event horizon and hence a black hole solution. For more discussion see text.}
\label{F3}
\end{center}
\end{figure}
The object exhibits a 5 GHz radio-flux density of $0.265$Jy. Using Lorentz factors $\Gamma=2$ and $\Gamma=5$, the emitted jet powers are calculated and reported in \ref{Table2} \cite{Pei:2016kka,Middleton:2014cha}. As before, an error of $0.3  ~dex$ is associated with the reported jet powers \cite{Narayan:2011eb,Middleton:2014cha}.
The emitted jet powers (corresponding to $\Gamma=2$ and $\Gamma=5$) along with their errors is compared with the theoretical jet power and the results are presented in \ref{f3a} and \ref{f3b} respectively. The values of $r_2$ and $a$ that can explain the emitted jet power within the error bars are denoted by the orange shaded region. The blue shaded region on the other hand, illustrates the allowed values of $r_2$ and $a$ when the theoretical radiative efficiency equals the observed $\eta$. As before $r_2>1.6$ cannot explain the observed $\eta$ while no such restriction on $r_2$ is imposed from the observed jet power. Once again, the maximum allowed magnitude of $r_2$ from both the observations is $r_2\sim 0.9$ (in both \ref{f3a} and \ref{f3b}). The range of spin predicted from the jet power (when $r_2=0$) is consistent with the range estimated by the Continuum Fitting method.

\item {\bf GRS 1124-683:} This X-ray binary comprises of a black hole of mass $11.0^{+2.1}_{-1.4} M_\odot$ \cite{Wu:2015ioy} and a K-type main sequence star as the companion with an orbital period of 10.4 hours \cite{1992ApJ...399L.145R}. The distance to the source is $D=4.95^{+0.69}_{-0.65}$ kpc while the inclination is $i = {43.2 ^{+2.1}_{-2.7}}^\circ$ \cite{Wu:2015ioy}. The spin of the object has been estimated by the Continuum Fitting method which turns out to be $a=0.63^{+0.16}_{-0.19}$ \cite {Chen:2015mvc}. Based on this value for the Kerr parameter, the radiative efficiency has been estimated (\ref{Table1}). The allowed values of $r_2$ and $a$ from observed $\eta$ are described by the blue shaded region in \ref{F4} which reveals that $r_{2,{max}}\sim 1.8$.

\begin{figure}[t!]
\begin{center}
\subfloat[\label{f4a}]{\includegraphics[scale=0.45]{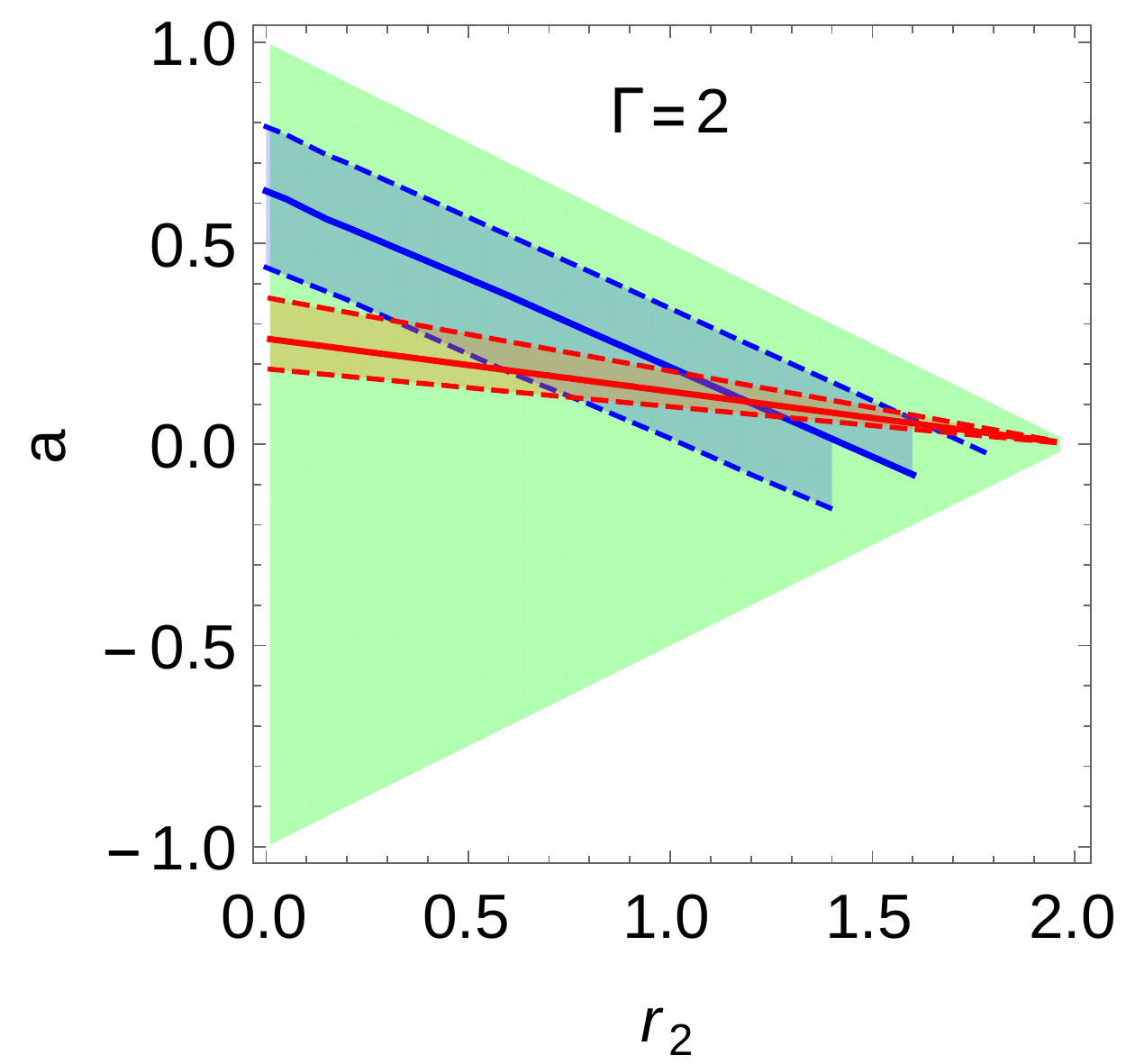}}
\hspace{0.5cm}
\subfloat[\label{f4b}]{\includegraphics[scale=0.45]{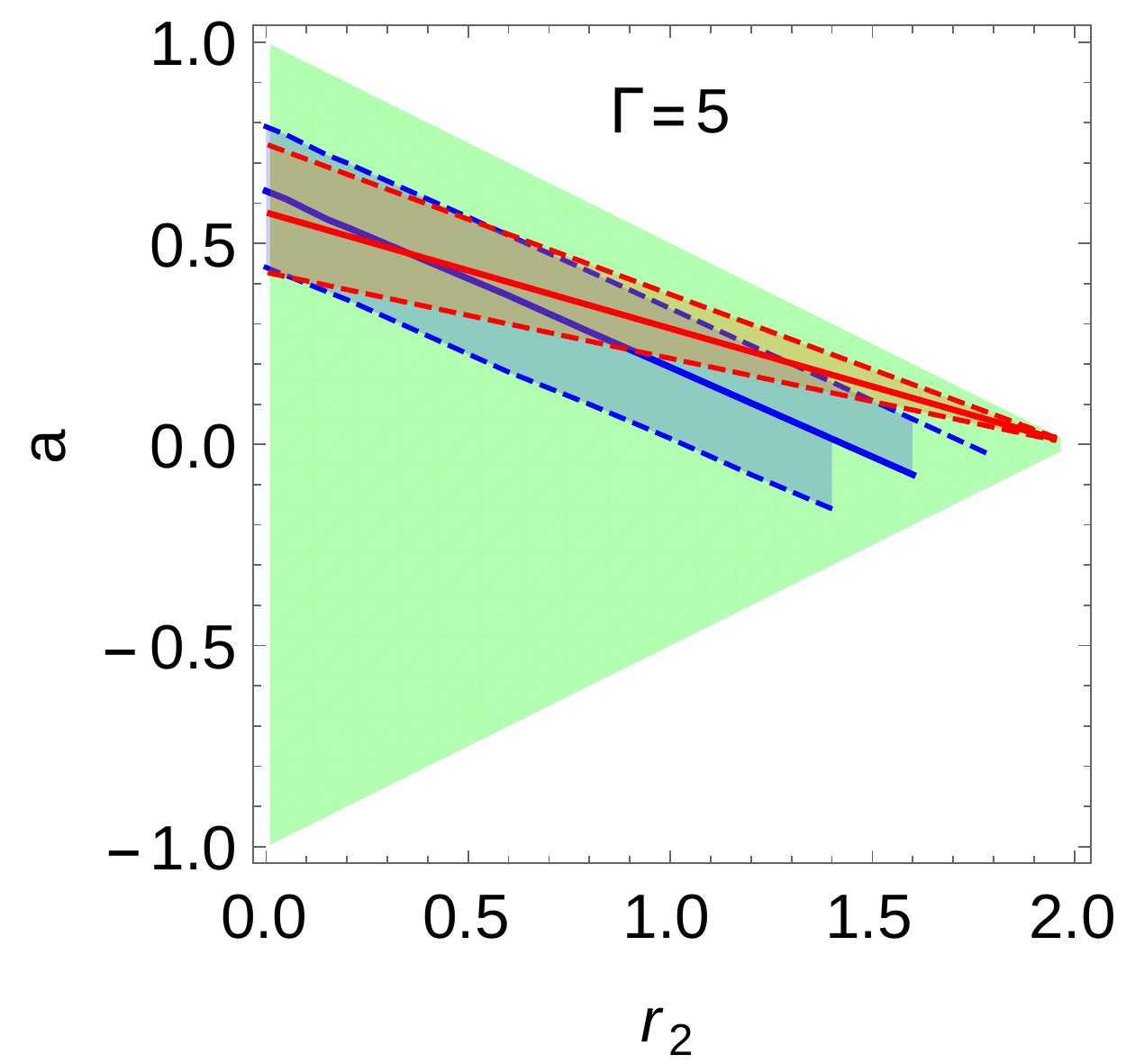}}
\caption{Black hole source GRS 1124-683: The description of the blue, orange and the green shaded regions remain the same as in the previous figures. The solid and dashed, red and blue lines also retain the same definition as in \ref{F1}.}
\label{F4}
\end{center}
\end{figure}

The emitted jet power corresponding to this source for $\Gamma=2$ and $\Gamma=5$ are reported in \ref{Table2}. A $0.3  ~dex$ error on the reported jet power is assumed \cite{Middleton:2014cha,Narayan:2011eb}. In \ref{F4} the orange shaded region represents the allowed values of $r_2$ and $a$ which can explain the emitted jet power within the error bars. 
We note that the maximum magnitude of $r_2$ that can explain both the observed $\eta$ and $P_{jet}$ corresponds to $r_2=1.7$ and $r_2=1.5$ for $\Gamma=2$ and $\Gamma=5$ respectively.
Moreover, unlike the previous black holes, the allowed range of spin that can describe both the observations when $r_2=0$, shows an overlap only when $\Gamma=5$ is considered to compute the emitted jet power from the observed 5 GHz radio-flux density.


\item {\bf GRO J1655-40:} GRO J1655-40 consists of a black hole of dynamical mass $M=6.3\pm 0.5 M_\odot$ \cite{Greene:2001wd} and an F-type secondary star of mass $M_S=2.34\pm 0.12 M_\odot$ with an orbital period of $2.62$ days \cite{Orosz:1996cg}. The  distance of the source has been estimated to be $D=3.2\pm 0.5$ kpc \cite{Hjellming:1995tv} while its orbital inclination turns out to be $i=70.2\pm 1.9 ^{\circ}$ \cite{Greene:2001wd}. There is a lot of controversy regarding the spin of this source. While the Continuum Fitting method predicts a spin $a\sim 0.65-0.75$ \cite{Shafee:2005ef}, the spin estimated by the Fe-line method is $a>0.9$ \cite{10.1111/j.1365-2966.2009.14622.x}. Based on the quasi-periodic oscillatons observed in the power spectrum of GRO J1655-40, the mass and spin of this object has been constrained to be $M=5.31\pm 0.07 M_\odot$ and $a=0.290 \pm 0.003$ respectively \cite{Motta:2013wga}. 
In this work however, we consider the spin estimated by the Continuum Fitting method to evaluate the radiative efficiency. As before, the allowed values of $r_2$ and $a$ that can explain the observed $\eta$ within the error bars are shaded in blue in \ref{F5} which shows that $r_{2,{max}}\sim 1.8$.  

\begin{figure}[t!]
\begin{center}
\subfloat[\label{f5a}]{\includegraphics[scale=0.45]{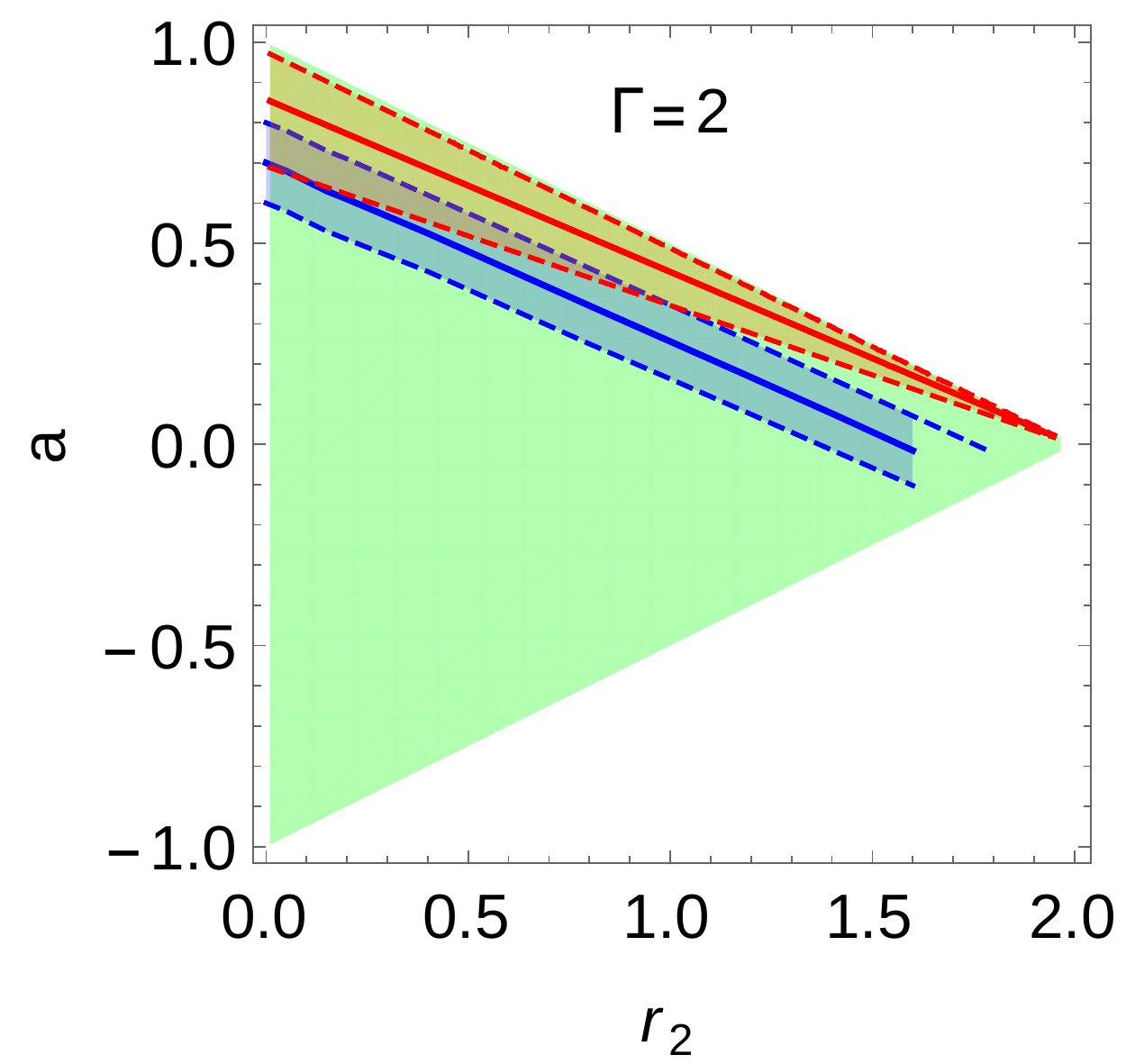}}
\hspace{0.5cm}
\subfloat[\label{f5b}]{\includegraphics[scale=0.45]{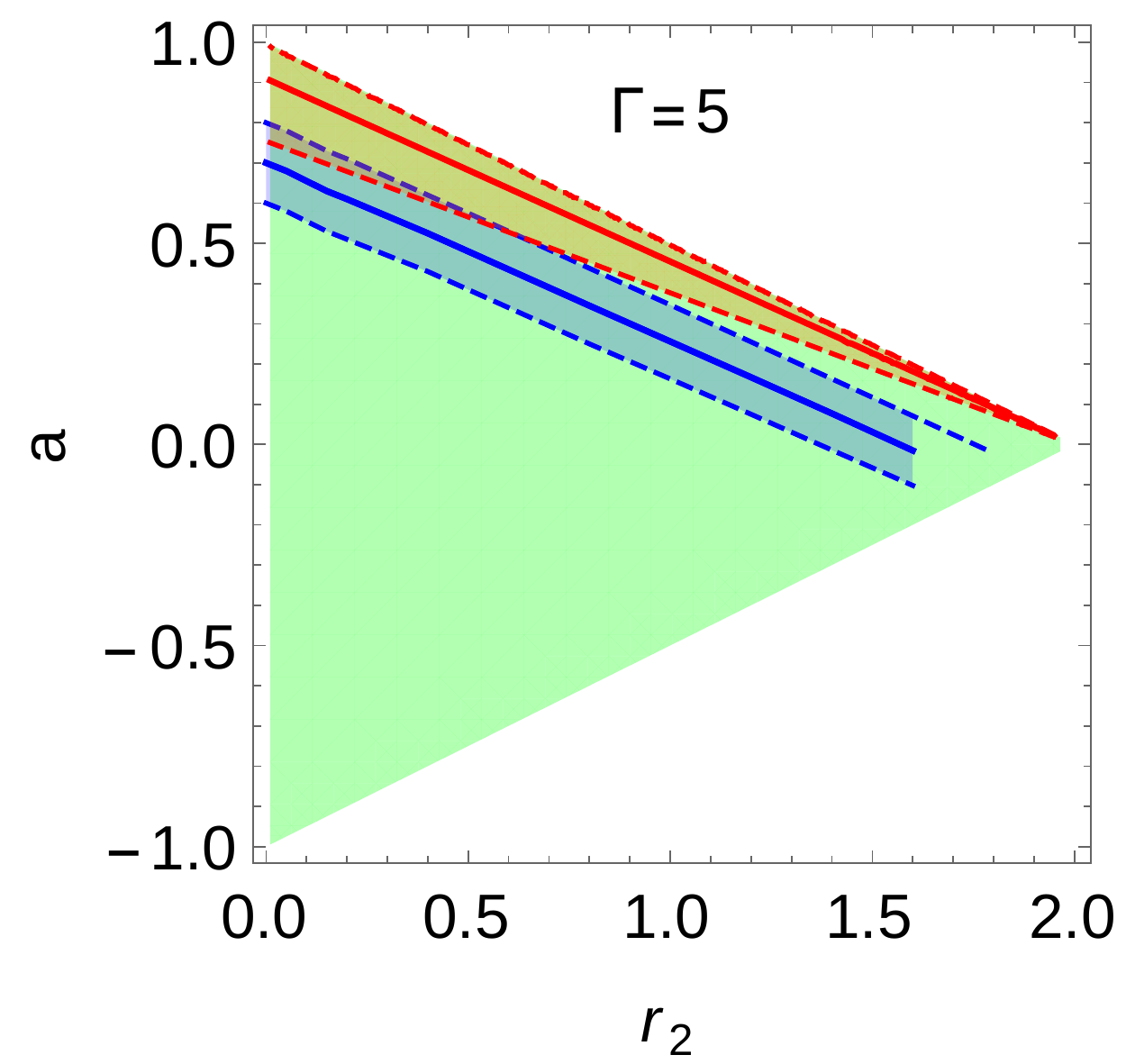}}
\caption{Black hole source GRO J1655-40: The description of the blue, orange and the green shaded regions remain the same as in the previous figures. The solid and dashed, red and blue lines also retain the same definition as in \ref{F1}.}
\label{F5}
\end{center}
\end{figure}
Using the 5 GHz radio-flux density of $2.42$Jy, the emitted jet power has been evaluated assuming Lorentz factors $\Gamma=2$ and $\Gamma=5$, which are reported in \ref{Table2}. These are associated with an error of $0.3  ~dex$. In \ref{F5} the orange shaded region represents the values of $r_2$ and $a$ that can explain the emitted jet power within the allowed errors. We note that \ref{f5a} and \ref{f5b} corresponds to the emitted jet power being computed using $\Gamma=2$ and $\Gamma=5$ respectively. We note that the observed jet power can be explained by the entire range of $r_2$. However, if we consider both the observations, $0\lesssim r_2\lesssim 1$ if $\Gamma=2$ and $0\lesssim r_2\lesssim 0.7$ when $\Gamma=5$ is assumed.

\item {\bf GRS 1915+105:} GRS 1915+105 is a galactic X-ray binary consisting of a  black hole and a K-type star with an orbital period of 34 days \cite{Greiner:2001vb,Mirabel:1994rb}.  
The black hole in this X-ray binary has a mass $M=12.4^{+2.0}_{-1.8}$ $M_\odot$ \cite{Reid:2014ywa}. The distance to the source is $8.6^{+2.0}_{-1.6}$ kpc and the inclination angle is $60\pm 5^\circ$ \cite{Reid:2014ywa}. The spin of the black hole estimated by the Continuum-Fitting method turns out to be $a>0.98$ \cite{McClintock:2006xd} which is used to estimate the radiative efficiency. In \ref{F6} the blue shaded region bounded by the blue dashed and solid line shows the allowed values of $r_2$ and $a$ that can explain the observed $\eta$. We note that a higher value of $r_2$ requires a lower spin to reproduce the radiative efficiency.

\begin{figure}[t!]
\begin{center}
\subfloat[\label{f6a}]{\includegraphics[scale=0.45]{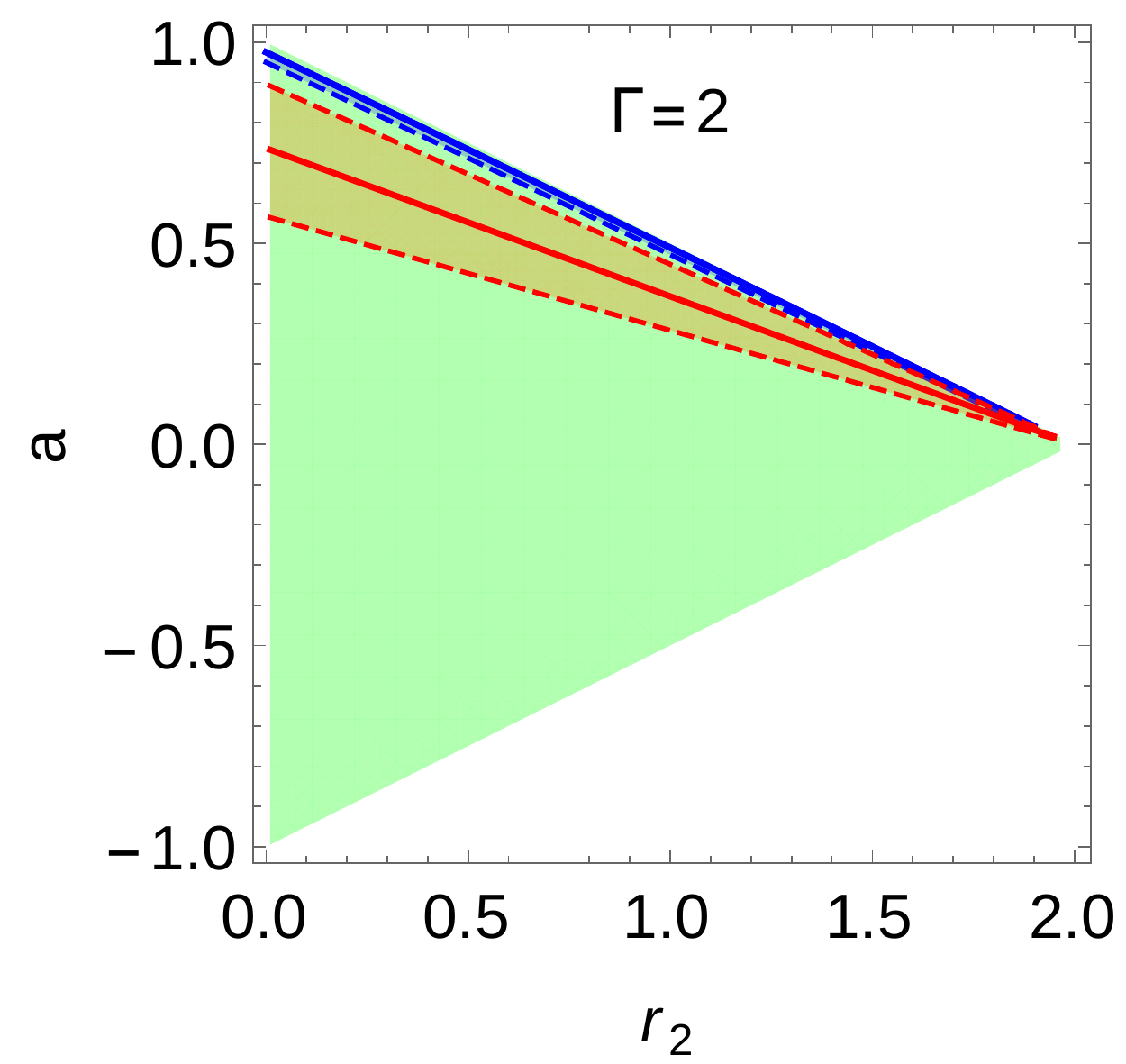}}
\hspace{0.5cm}
\subfloat[\label{f6b}]{\includegraphics[scale=0.45]{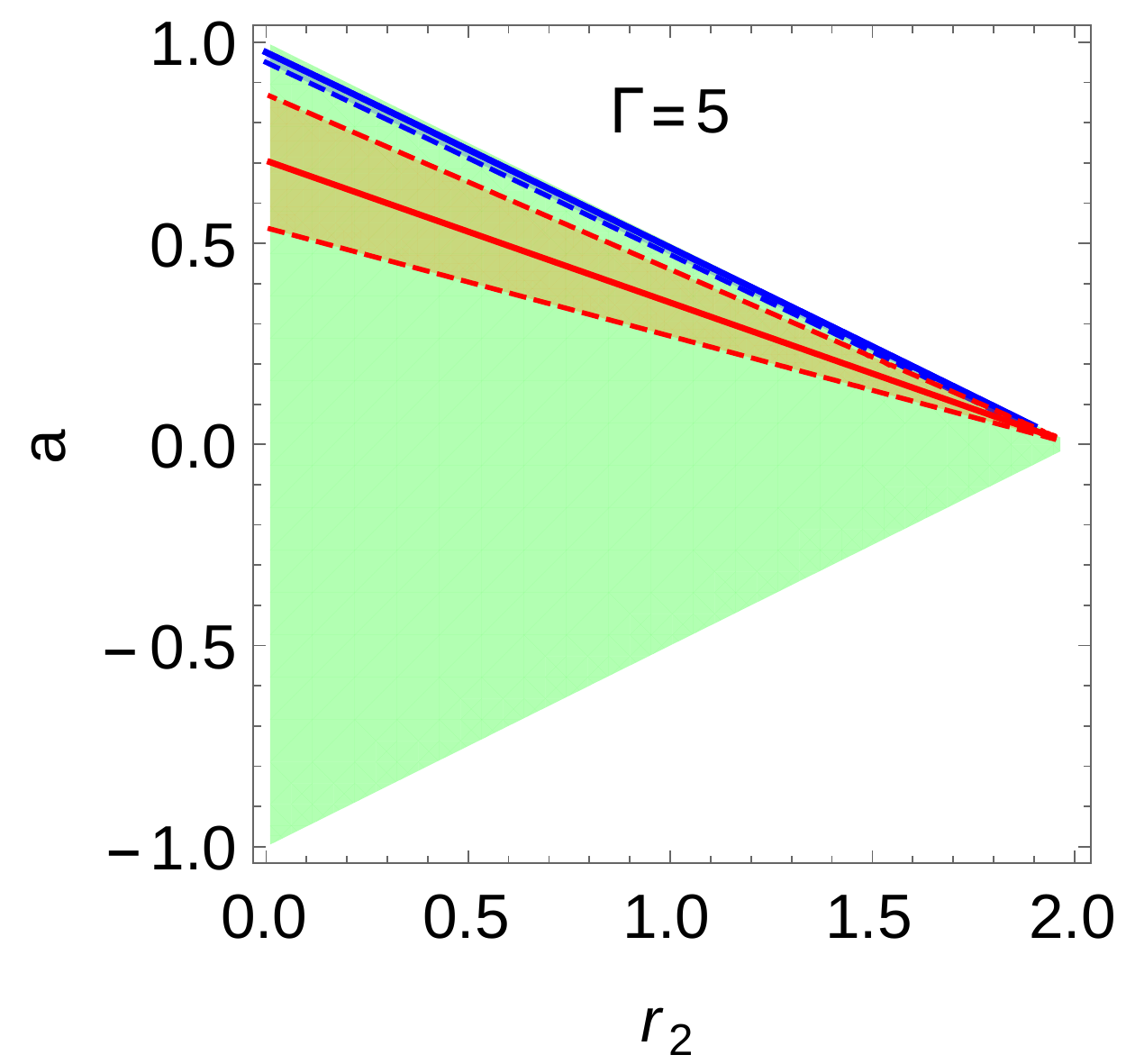}}
\caption{Black hole source GRS 1915+105: The description of the blue, orange and the green shaded regions remain the same as in the previous figures. The solid and dashed, red and blue lines also retain the same definition as in \ref{F1}.}
\label{F6}
\end{center}
\end{figure}

The object exhibits strong radio-jets with 5 GHz radio-flux density being 0.912 Jy \cite{Mirabel:1994rb}. The emitted jet power derived from the flux density after Doppler de-boosting with Lorentz factors $\Gamma=2$ and $\Gamma=5$ are reported in \ref{Table2}. As before the error associated with the jet power is $0.3 ~dex$. The orange shaded region in \ref{F6} indicates the values of $r_2$ and $a$ which can address the observed jet power within the error bars. The solid and the dashed red lines bear the same definition as before. We note from \ref{F6} that almost the entire allowed range of $r_2$ can explain both the observed $P_{jet}$ and $\eta$. Also, when $r_2 \gtrsim 1.5$ both the observations can be simulataneously explained.

\end{itemize}

\subsection{Implications on the axion-dilaton parameters from observational constraints}
We have noted in the last section that the observed radiative efficiency and the jet power can be used to discern the observationally favored magnitude of the dilaton parameter. In order to gain a better understanding on this we note that comparison of the theoretical radiative efficiency (\ref{S3-5}) and the jet power (\ref{6}) with the corresponding observations of six microquasars (as depicted in \ref{F1} to \ref{F6}) exhibit a few common features:
\begin{itemize}
\item The observed jet power can be explained by almost the entire allowed range of $r_2$.
\item A higher value of $r_2$ requires a lower $a$ to explain the observed $P_{jet}$ and $\eta$.
\item The observational bound on $r_2$ arises when one tries to reproduce the observed $\eta$.
\item In most of the cases, when $r_2=0$ (general relativistic scenario), the observationally allowed range of $a$ obtained from $P_{jet}$ and $\eta$ exhibit an overlap.  
\end{itemize} 
The above features motivate us to evaluate the chi-square as a function of $r_2$ by comparing $P_{BZ}$ (\ref{6}) and $\eta$ (\ref{S3-5}) with the corresponding observations. 
This corresponds to the joint-$\chi^2$ given by,
\begin{align}
\chi ^2 (r_2,\lbrace a \rbrace)=\sum_{i} \frac{\lbrace \eta_{obs{,i}}-\eta(r_2,\lbrace a \rbrace) \rbrace ^2}{\sigma_{\eta, i}^2}  +  \sum_{i}  \frac{\lbrace P_{jet} -P_{BZ}  (r_2,\lbrace a \rbrace) \rbrace ^2}{\sigma_{P, i}^2}
\label{10}
\end{align}
For every $r_2$ we vary $a$ in the allowed range: $-(1-\frac{r_2}{2})\leq a \leq (1-\frac{r_2}{2})$ (such that the event horizon exists) and compute $\chi ^2 (r_2,\lbrace a \rbrace)$ as in \ref{10}. The spin parameter which gives the minimum $\chi^2$ for the chosen $r_2$, is considered to be the $\chi^2$ for that $r_2$. Repeating this procedure for all values of $r_2$ in the range $0\leq r_2 \leq 2$, we obtain the variation of $\chi^2$ with $r_2$. 

\begin{figure}[t!]
\centering
\includegraphics[scale=0.75]{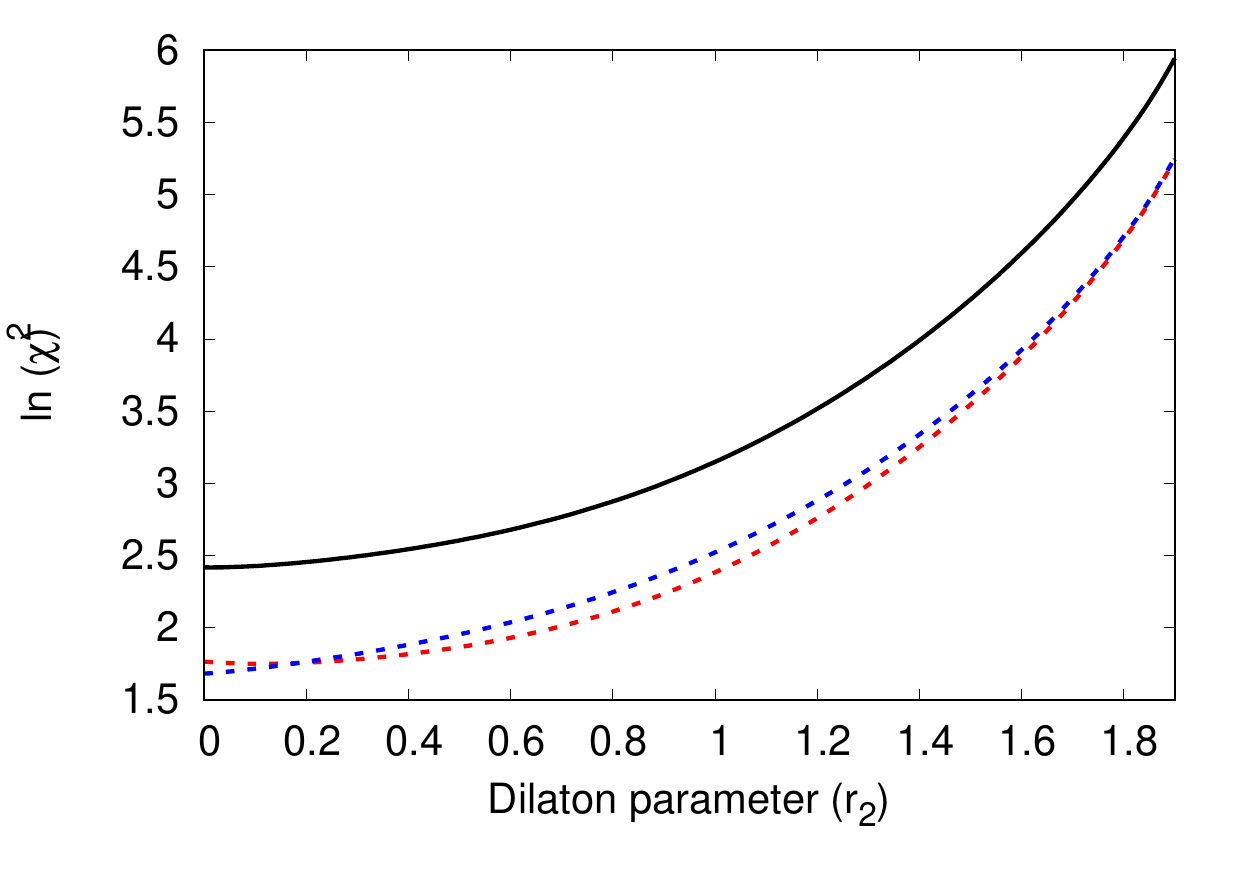}
\caption{The figure illustrates the variation of $\chi^{2}$ with the dilaton parameter $r_2$ for the sample of microquasars. The red dashed line represents the situation when $P_{jet}$ corresponding to $\Gamma=2$ is used to compute the $\chi^{2}$, while the blue dashed line is associated with the scenario when $P_{jet}$ corresponding to $\Gamma=5$ is considered for evaluating the $\chi^{2}$. The black solid line denotes the joint-$\chi^2$ when both $\Gamma=2$ and $\Gamma=5 $ are taken into account.
For more discussions see text.}
\label{F7}
\end{figure}

\ref{F7} shows the variation of the natural logarithm of the $\chi^2$ computed by the above procedure with the dilaton parameter $r_2$. The red and blue dashed lines are associated with the situation when $P_{BZ}$ in \ref{10} is compared with the observed $P_{jet}$ corresponding to $\Gamma=2$ and $\Gamma=5 $, respectively. The black solid line denotes the joint-$\chi^2$ when both $\Gamma=2$ and $\Gamma=5 $ are taken into account. From \ref{F7}, we note that while the $\Gamma=2$ scenario slightly favors a non-zero dilaton parameter ($r_2\sim 0.1$), the $\Gamma=5 $ case favors the general relativistic scenario. From the joint-$\chi^2$, the observationally favored dilaton parameter seems to be $r_2\sim 0$. The most important outcome of this analysis is that the extreme or even moderate values of $r_2$ are disfavored from observations related to jet power and radiative efficiency. This implies that pure dilaton black holes (which are non-rotating) are less favored compared to their axion-dilaton counterparts. Also, since the joint-$\chi^2$ minimizes around $r_2\sim 0$, the Kerr black holes seem to explain the observations better than Kerr-Sen black holes.

\section{Summary and concluding remarks}
\label{Sec5}
In this work we aim to discern the imprints of Einstein-Maxwell dilaton-axion gravity from observations related to jet power and radiative efficiency of microquasars.
The EMDA gravity essentially arises in the low energy effective action of superstring theories and investigating the observational signatures of the same is important as it can provide an indirect testbed for string theory. The theoretical implications of this model has been explored extensively in the past and the exact, stationary and axi-symmetric black hole solution in this theory has been worked out. Such a solution corresponds to the Kerr-Sen spacetime which contains dilaton charges while the axionic field renders angular momentum to such black holes. 

The observational signatures of the Kerr-Sen spacetime has been explored in the context of strong gravitational lensing and black hole shadows \cite{Gyulchev:2006zg,An:2017hby,Younsi:2016azx,Hioki:2008zw}. Therefore, in this work we investigate the role of the Kerr-Sen background in affecting the jet power and the radiative efficiency derived from the continuum spectrum associated with the black holes.
The transient jet power and the peak emission of the continnum spectrum from the accretion disk are sensitive to the background spacetime and hence can be used as important observational tools to probe the nature of strong gravity.

The power associated with transient jets is computed based on the Blandford-Zanjeck model which explicitly reveals the dependence of the background metric on the jet power. This is then compared with the emitted jet power of a sample of microquasars estimated from the peak 5 GHz radio flux density which is Doppler boosted by Lorentz factors 
$\Gamma=2$ and $\Gamma=5$ and scaled by the distance to obtain the associated luminosity.  
The jet power estimated by this method turns out to be model independent and hence sufficiently reliable \cite{Narayan:2011eb}. 
If the background is governed by the Kerr metric then such an observation can be used to determine the black hole spins \cite{Narayan:2011eb}. In the event the background corresponds to the Kerr-Sen spacetime, the allowed values of the spin and the dilaton parameters can be determined based on the observed jet power.
A departure from \gr\ therefore introduces a degeneracy between the metric parameters and only a combination of these parameters can be constrained.  

The radiative efficiency, the second metric dependent quantity used in this work, is calculated based on the Novikov-Thorne model for thin accretion disk. This is subsequently compared with the observed radiative efficiency of the same sample of microquasars whose jet powers have been evaluated. The radiative efficiency is derived from the peak emission of the continuum spectrum and assuming \gr, it can be used to constrain the spins of the underlying black holes. This forms the basis of the Continuum-Fitting method for determining the spins of the microquasars. In the event the background is governed by the Kerr-Sen spacetime, the radiative efficiency can be used to determine the allowed values of spin and dilaton parameters for each of the black holes. 

We note that the dilaton-axion black hole can explain the observed jet power and the radiative efficiency of the microquasars. Although, in this case the jet is powered by the interplay between the axion and the dilaton fields.
For each of the microquasars the spin and the dilaton parameters which can explain both the observations are considered. It turns out that in most of the cases large values of the dilaton parameters, viz, $r_2\gtrsim 1$ are generally disfavored. A greater axionic field strength requires a smaller dilatonic charge of the black hole to reproduce these two observations. A chi-square analysis is performed where the observed jet power and the radiative efficiency of the microquasars are compared with the corresponding theoretical estimates depending on the metric parameters. Such an analysis clearly reveals that pure dilaton black holes are observationally less favored compared to their axion-dilaton counterparts. Moreover, since the chi-square minimizes when $r_2\simeq 0$, the Kerr black holes seem to be observationally more favored compared to the Kerr-Sen black holes. We have noted earlier that $r_2/\mathcal{M}=\frac{\alpha^\prime Q^2}{8\mathcal{M}^2}$. Therefore obtaining $r_2\simeq 0$ from the observations implies $Q\simeq 0$ since $\alpha^\prime \simeq 0$ would lead to gauge anomaly. Since astrophysical black holes are expected to carry negligible charge \cite{Blandford:1977ds}, our result $r_2\simeq 0$ also implies $Q\simeq 0$, in which case we retrieve the Kerr metric. However, it is important to note that observational validation of the Kerr scenario does not necessarily validate general relativity since the Kerr metric also arises as black hole solution for several alternative gravity scenarios \cite{Sen:1992ua,Campbell:1992hc,Psaltis:2007cw}.

In the Kerr scenario, the axion or the Kalb-Ramond field exhibit a vanishing field strength whose suppression has been observed in several other physical scenarios, e.g in the context of higher curvature gravity where the related scalar degrees of freedom reduces the coupling of such a field with the Standard Model fermions \cite{Paul:2018ycm,Das:2018jey}, in the warped braneworld scenario \cite{Randall:1999ee} with bulk Kalb-Ramond fields \cite{Mukhopadhyaya:2001fc,Mukhopadhyaya:2002jn} and the related stabilization of the modulus \cite{Das:2014asa} and in the inflationary era induced by higher curvature gravity \cite{Elizalde:2018rmz,Elizalde:2018now} and higher dimensions \cite{Paul:2018jpq}. A similar conclusion $r_2\simeq 0.2$ (which is close to $r_2\simeq 0$) has been independently achieved by comparing the theoretical luminosity from the accretion disk in the Kerr-Sen background with the optical observations of quasars \cite{Banerjee:2020qmi}.
This result can be further verified with the availability of a larger observational sample or by considering more observations in the electromagnetic domain, e.g. quasi-periodic oscillations or black hole shadow, which will be reported in a future work.

\section*{Acknowledgements}
The research of SSG is supported by the Science and Engineering Research Board-Extra Mural Research Grant No. (EMR/2017/001372), Government of India.
\appendix

\labelformat{section}{Appendix #1}
\labelformat{subsection}{Appendix #1}
\labelformat{subsubsection}{Appendix #1}

\section*{Appendix}
\section{Derivation of the jet power in the Blandford-Znajeck model}
\label{AA}

In this section we derive the jet power in the Blandford-Znajeck model assuming a general stationary, axi-symmetric spacetime. We have already discussed that the Blandford-Znajeck model assumes a force-free magnetosphere where the particle inertia is neglected compared to the the energy-momentum tensor due to the electromagnetic fields, such that,
\begin{align}
\label{A1}
T_{\mu\nu}^{tot}=T_{\mu\nu}^{EM}+T_{\mu\nu}^{matter}\approx T_{\mu\nu}^{EM} \tag{A1}
\end{align}  
where,
\begin{align}
\label{A2}
T_{\mu\nu}^{EM}=F_{\mu\rho} F^\rho_\nu- \frac{1}{4} g_{\mu\nu}F^{\alpha\beta}F_{\alpha\beta} \tag{A2}
\end{align} 
satisfies the conservation equation,
\begin{align}
\label{A3}
\nabla^\mu T_{\mu\nu}^{EM}=0
\tag{A3}
\end{align}
In \ref{A2}, $F_{\mu\nu}=\partial_\mu A_\nu-\partial_\nu A_\mu$ is the Faraday tensor and $A_\mu$ is the gauge field. \\
In a force-free magnetosphere the Faraday tensor satisfies the relation \cite{Blandford:1977ds},
\begin{equation}
\label{A4}
F_{\mu\nu}J^\nu=0
\tag{A4}
\end{equation}
such that 
\begin{subequations}
\begin{equation}
A_{t,r}J^r+A_{t,\theta}J^\theta=0 \tag{A4a}
\label{4a}
\end{equation}
\begin{equation}
A_{t,r}J^t+A_{\phi,r}J^\phi+B_\phi J^\theta=0 \tag{A4b}
\label{4b}
\end{equation}
\begin{equation}
A_{t,\theta}J^t+A_{\phi,\theta}J^\phi+B_\phi J^r=0   \tag{A4c}
\label{4c}
\end{equation}
\begin{equation}
A_{\phi,r}J^r+A_{\phi,\theta}J^\theta=0   \tag{A4d}
\label{4d}
\end{equation}
\end{subequations}
and $J^\nu$ is the current 4-vector.
From \ref{4a} and \ref{4d} one can define the electromagnetic angular velocity $\omega(r,\theta)$, where,
\begin{align}
\label{A5}
\frac{A_{t,r}}{A_{\phi,r}}=\frac{A_{t,\theta}}{A_{\phi,\theta}}=-\omega(r,\theta) \tag{A5}
\end{align}
Assuming the validity of the force-free condition and that $A_\mu$ is stationary and axi-symmetric one can write the Faraday tensor in the form,
\begin{equation}
\label{A6}
F_{\mu\nu}=\sqrt{-g}
\begin{pmatrix}
0 & -\omega B^\theta & \omega B^r & 0 \\
\omega B^\theta & 0 & B^\phi & -B^\theta \\
-\omega B^r & -B^\phi & 0 & B^r \\
0 & B^\theta & -B^r & 0 
\end{pmatrix}  \tag{A6}
\end{equation}

The power associated with the relativistic jet in the context of the Blandford-Znajeck model is given by,
\begin{align}
\label{A7}
P_{BZ}=4\pi \int_0^{\pi/2} \sqrt{-g}~T^r_t~d\theta  \tag{A7}
\end{align}
which takes into account the fact that the jets are bipolar. In \ref{A7} $T^r_t$ represents the radial component of the Poynting flux.
From \ref{A2} it can be shown that the radial component of the Poynting flux assumes the form,
\begin{align}
\label{A8}
T^r_t=g^{rr}g^{\theta\theta}F_{r\theta}F_{\theta t}-g^{rt}g^{\theta\theta}F_{t\theta}^2+g^{r\phi}g^{\theta\theta}F_{\phi\theta}F_{\theta t} \tag{A8}
\end{align}
such that the information of the metric enters both through its determinant and through $T^r_t$ in the jet power.

\section{Jet power in the Einstein-Maxwell-dilaton-axion gravity}
\label{AB}
In this section we derive the jet power in the Kerr-Sen background arising in Einstein-Maxwell-dilaton-axion gravity.
We assume that the jet launching radius corresponds to the event horizon and hence the first term in \ref{A8} vanishes. In order to derive the jet power one requires that the metric is regular at the horizon. As a result we express our metric given by \ref{S2-11} in the Kerr-Schild coordinates,

\begin{align}
\label{B1}
ds^2= \bigg(-1+\frac{2r}{\tilde{\Sigma}} \bigg)dt^2+\frac{4r}{\tilde{\Sigma}}dt dr- \mathrm{sin}^2\theta \frac{4ar}{\tilde{\Sigma}}dtd\phi + \bigg(1+\frac{2r}{\tilde{\Sigma}} \bigg)dr^2- 2a\mathrm{sin}^2\theta \bigg(1+\frac{2r}{\tilde{\Sigma}} \bigg)dr d\phi +\tilde{\Sigma}d\theta^2+ & \nonumber \\
\mathrm{sin}^2\theta\bigg(a^2+r(r+r_2)+2\mathrm{sin}^2\theta\frac{a^2r}{\tilde{\Sigma}}\bigg)d\phi^2&  \tag{B1}
\end{align}
Using \ref{A8} and \ref{B1} the radial component of the Poynting flux is given by,
\begin{align}
\label{B2}
T^r_t=2r_H M sin^2\theta(B^r)^2\omega \big[\Omega_H-\omega\big]\biggr\rvert_{r=r_{H}} \tag{B2}
\end{align}
where, 
\begin{align}
\label{B3}
r_H=1-\frac{r_2}{2}+\sqrt{\bigg(1-\frac{r_2}{2}\bigg)^2-a^2} \tag{B3}~~\rm and 
\end{align} 
\begin{align}
\label{B4}
\Omega_H=\bigg(-\frac{g_{t\phi}}{g_{\phi\phi}}\bigg)\biggr\rvert_{r=r_H} \tag{B4}
\end{align}
is the horizon radius and the angular velocity of the event horizon respectively.  
It is important to note that for the metric in \ref{B1}, 
\begin{align}
\label{B5}
\Omega_H=\bigg(-\frac{g_{t\phi}}{g_{\phi\phi}}\bigg)\biggr\rvert_{r=r_H}=\bigg(\frac{g^{r\phi}}{g^{rt}}\bigg)\biggr\rvert_{r=r_H}=\frac{a}{2r_H}  \tag{B5}
\end{align}
and $\sqrt{-g}=\tilde{\Sigma}sin\theta$.

At this stage, it is impossible to calculate the jet power without knowing the form of $\omega$ and $B^r$. This requires solving \ref{A3} which is quite non-trivial. Therefore, we follow the approach adopted by \cite{Tchekhovskoy:2009ba,Tanabe:2008wm}, where $\omega=\Omega_H/2$ is assumed. This can be obtained by maximizing the radial Poynting flux $T^r_t$ in \ref{B2} with respect to $\omega$ \cite{Tchekhovskoy:2009ba}. With this assumption, \ref{B2} is given by,
\begin{align}
\label{B6}
T^r_t=2r_H M sin^2\theta(B^r)^2\frac{\Omega_H^2}{4} \tag{B6}
\end{align}
In the stationary and axi-symmetric spacetime at a constant $(r,\theta)$, the physical quantities are invariant along the azimuthal direction, the so called `m-loops' \cite{MacDonald:1982zz}. Consequently, by applying Stoke's law along one of these `m-loops' the magnetic flux $\Phi_B$ through it is given by, 
\begin{align}
\label{B7}
\Phi_B=\int \vec{B}.\vec{dS}=\int (\vec{\nabla} \times \vec{A}).\vec{dS}=\oint \vec{A}.\vec{dl}=2\pi A_\phi  \tag{B7}
\end{align}
Further, from \ref{A6} it is clear that $F_{\theta\phi}= A_{\phi,\theta}=\sqrt{-g} B^r $ such that,
\begin{align}
\label{B8}
\Phi_B=2\pi A_\phi=2\pi \int_0^\pi \sqrt{-g} |B^r| d\theta=2\pi \Psi  \tag{B8}
\end{align}
Therefore, the azimuthal component of the vector potential $A_\phi$ is directly related to the magnetic flux through the m-loops and is denoted by $\Psi$.
Assuming that the magnetic flux $\Psi$ approximately follows the split-monopole profile it can be shown that \cite{Tchekhovskoy:2009ba},
\begin{align}
\label{B9}
B^r=\frac{\Psi_{tot}}{r^2} \tag{B9}
\end{align} 
where terms of the order $\Omega_H^2$ and higher are neglected.
Substituting \ref{B9} and \ref{B6} in \ref{A7} and evaluating it at the horizon radius, it can be shown that
\begin{align}
\label{B10}
P_{BZ}=k\Phi_{tot}^2\Omega_H^2 \tag{B10}
\end{align}
when terms of order $\Omega_H^4$ and higher are neglected \cite{Tchekhovskoy:2009ba}. In \ref{B10} $\Phi_{tot}$ represents the magnetic flux threading the event horizon.

\label{S5}

\bibliography{KN-ED,Brane,Black_Hole_Shadow,EMDA-Jet,Gravity_1_full,Gravity_2_full,Gravity_3_partial,My_References,extra-references}
\bibliographystyle{utphys1}
\end{document}